\pdfoutput=1 % arXiv: force pdfLaTeX (PDF/PNG figures); must stay in the first 5 lines
\documentclass[sigconf]{acmart}

\usepackage{xcolor}
\usepackage{listings}
\usepackage{booktabs}
\usepackage{multirow}
\usepackage{makecell}
\usepackage{enumitem}
\usepackage{subcaption}
\usepackage{latexml}
\providecommand{\Description}[2][]{}
\providecommand{\setcctype}[2][]{}
\AtBeginDocument{%
  }

\newcommand{\sys}{MAESTRO}
\definecolor{maestroblue}{HTML}{119EC9}
\newcommand{\circnum}[1]{%
  \iflatexml\textcircled{#1}%
  \else{\Large\textcircled{\normalsize\textbf{#1}}}\fi}
\newcommand{\circnumblue}[1]{%
  \iflatexml\textcolor{maestroblue}{\textcircled{#1}}%
  \else{\textcolor{maestroblue}{\Large\textcircled{\normalsize\textbf{#1}}}}\fi}

\lstdefinestyle{json}{
  basicstyle=\ttfamily\small,
  columns=fullflexible,
  showstringspaces=false,
  breaklines=true,
  frame=single,
  rulecolor=\color{gray!40},
}

\newcommand{\rev}[1]{#1}
\newenvironment{revblock}{}{}

\copyrightyear{2026}
\acmYear{2026}
\setcopyright{cc}
\setcctype{by}
\acmConference[UIST '26]{The 39th Annual ACM Symposium on User Interface Software and Technology}{November 02--05, 2026}{Detroit, MI, USA}
\acmBooktitle{The 39th Annual ACM Symposium on User Interface Software and Technology (UIST '26), November 02--05, 2026, Detroit, MI, USA}
\acmDOI{10.1145/3830398.3830672}
\acmISBN{979-8-4007-2856-3/2026/11}

\begin{document}

\title{\sys{}: Adapting GUIs and Guiding Navigation with User Preferences in Conversational Agents with GUIs}

\author{Sangwook Lee}
\orcid{0000-0002-2600-4769}
\affiliation{%
  \department{Computer Science}
  \institution{Virginia Tech}
  \city{Blacksburg}
  \state{Virginia}
  \country{USA}
}
\email{sangwooklee@vt.edu}

\author{Adnan Abbas}
\orcid{0009-0005-8728-875X}
\affiliation{%
  \department{Computer Science}
  \institution{Virginia Tech}
  \city{Blacksburg}
  \state{Virginia}
  \country{USA}
}
\email{adnana99@vt.edu}

\author{Sang Won Lee}
\orcid{0000-0002-1026-315X}
\affiliation{%
  \department{Computer Science}
  \institution{Virginia Tech}
  \city{Blacksburg}
  \state{Virginia}
  \country{USA}
}
\affiliation{%
  \institution{NAVER AI Lab}
  \city{Seongnam}
  \country{Republic of Korea}
}
\email{sangwonlee@vt.edu}

\author{Young-Ho Kim}
\orcid{0000-0002-2681-2774}
\affiliation{%
  \institution{NAVER AI Lab}
  \city{Seongnam}
  \country{Republic of Korea}
}
\email{yghokim@younghokim.net}

\author{Yan Chen}
\orcid{0000-0002-1646-6935}
\affiliation{%
  \department{Computer Science}
  \institution{Virginia Tech}
  \city{Blacksburg}
  \state{Virginia}
  \country{USA}
}
\email{ych@vt.edu}

\begin{abstract}
Modern task-oriented chatbots present GUI elements alongside natural-language dialogue, yet the agent's role has largely been limited to interpreting natural-language input as GUI actions and following a linear workflow. In preference-driven, multi-step tasks such as booking a flight or reserving a restaurant, earlier choices constrain later options and may force users to restart from scratch. User preferences serve as the key criteria for these decisions, yet existing agents do not systematically leverage them. We present \sys{}, which extends the agent's role from execution to decision support. \sys{} maintains a shared preference memory that extracts hard and soft preferences from natural-language utterances and provides two mechanisms. \emph{Preference-Grounded GUI Adaptation} applies in-place operators (augment, sort, filter, and highlight) to the existing GUI according to preference strength, supporting comparison among options. \emph{Preference-Guided Workflow Navigation} detects conflicts between preferences and available options, proposes backtracking, and records failed paths to avoid revisiting dead ends. Through a controlled experiment ($N=33$), we demonstrated that \sys{} \rev{improved decision quality in movie ticketing: final bookings left fewer hard preferences unmet, and users made fewer selections that violated their stated preferences during the process than in the baseline condition, although task success rate and completion time did not differ significantly}. In addition, we showed that using \sys{} in voice mode can increase users' active engagement as well as their mental burden, revealing the nuanced tension in agentic interaction design for conversational agents with a GUI.
% tasks, without causing  \sys{} in a movie-booking Conversational Agent with GUI (CAG) through a 2 (Condition: Baseline vs.\ \sys{}) $\times$ 2 (Mode: Text vs.\ Voice) within-subjects study ($N=33$), 
\end{abstract}

% CCS concepts generated against the official ACM CCS 2012 taxonomy
% (concept ids verified from dl.acm.org/pb-assets/dl_ccs/acm_ccs2012 xml).
\begin{CCSXML}
<ccs2012>
   <concept>
       <concept_id>10003120.10003121.10003124.10010870</concept_id>
       <concept_desc>Human-centered computing~Natural language interfaces</concept_desc>
       <concept_significance>500</concept_significance>
   </concept>
   <concept>
       <concept_id>10003120.10003121.10003124.10010865</concept_id>
       <concept_desc>Human-centered computing~Graphical user interfaces</concept_desc>
       <concept_significance>500</concept_significance>
   </concept>
   <concept>
       <concept_id>10003120.10003121.10003129.10010885</concept_id>
       <concept_desc>Human-centered computing~User interface management systems</concept_desc>
       <concept_significance>300</concept_significance>
   </concept>
   <concept>
       <concept_id>10003120.10003121.10011748</concept_id>
       <concept_desc>Human-centered computing~Empirical studies in HCI</concept_desc>
       <concept_significance>300</concept_significance>
   </concept>
</ccs2012>
\end{CCSXML}

\ccsdesc[500]{Human-centered computing~Natural language interfaces}
\ccsdesc[500]{Human-centered computing~Graphical user interfaces}
\ccsdesc[300]{Human-centered computing~User interface management systems}
\ccsdesc[300]{Human-centered computing~Empirical studies in HCI}

\keywords{conversational agents with GUIs, adaptive user interfaces}

\begin{teaserfigure}
  \centering
  \includegraphics[width=\textwidth, trim=0 20 0 8, clip]{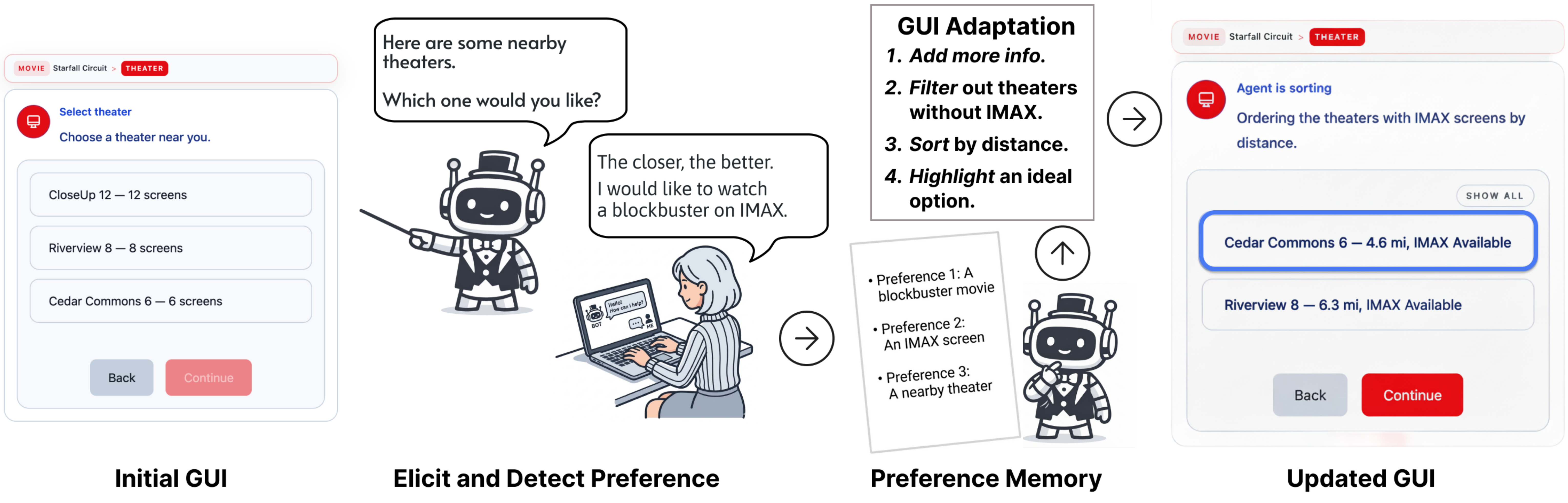}
  \caption{Overview of \sys{} at the theater selection stage. \textbf{Initial GUI:} The default GUI presents three theater options with names and screen counts. \textbf{Elicit and Detect Preference:} The agent asks about theater choice, and the user replies with preferences. \textbf{Preference Memory:} \sys{} extracts three preferences and plans four GUI adaptation operators. \textbf{Updated GUI:} The adapted GUI shows only IMAX-available theaters sorted by distance, enabling the user to compare and confirm directly.}
  \Description{A left-to-right diagram illustrating the \sys{} pipeline at the theater selection stage, consisting of four panels connected by arrows. The first panel, labeled Initial GUI, shows a movie-ticketing interface with the header ``Starfall Circuit'' and a ``THEATER'' tab. Three theater buttons are listed vertically: CloseUp 12 with 12 screens, Riverview 8 with 8 screens, and Cedar Commons 6 with 6 screens, along with Back and Continue buttons. The second panel, labeled Elicit and Detect Preference, depicts a robot agent character asking ``Here are some nearby theaters. Which one would you like?'' and an astronaut user character responding ``I would like to watch a blockbuster on an IMAX screen. The closer the better!'' The third panel, labeled Preference Memory, shows the robot agent holding a clipboard listing three extracted preferences: Preference 1 is a blockbuster movie, Preference 2 is an IMAX screen, and Preference 3 is a nearby theater. Above, a box titled GUI Adaptation enumerates four planned operators: 1 Add more info, 2 Filter out theaters without IMAX, 3 Sort by distance, and 4 Highlight an ideal option. The fourth panel, labeled Updated GUI, shows the adapted interface. A banner reads ``Agent is sorting --- Ordering the theaters with IMAX screens by distance.'' Below, only two theaters remain after filtering: Cedar Commons 6 at 4.6 miles with IMAX Available, highlighted with a red border, and Riverview 8 at 6.3 miles with IMAX Available. A ``SHOW ALL'' link appears above the filtered list, and Back and Continue buttons are at the bottom.}
  \label{fig:teaser}
\end{teaserfigure}

\maketitle

%% ============================================================
\section{Introduction}
\label{sec:introduction}
%% ============================================================

% Yan's version

Conversational Agents with GUIs (CAGs) are chatbots that interleave structured GUI widgets, such as buttons or drop-down menus, with natural-language dialogue.
They are proliferating across domains: personal banking (Erica~\cite{bankofamerica2024erica}), customer support (Xfinity~\cite{bankofamerica2024erica}), and recommendations (Amazon Rufus~\cite{chilimbi2024technology}, Booking.com~\cite{bookingcom2023aiplanner}). GUIs accelerate input, organize options for visual comparison, and guide intent through predefined choices~\cite{nguyen2022user}. Yet interacting with a GUI through natural language does not automatically unlock its full expressive potential. 

In current CAGs, the agent's role is largely limited to event handling, mapping GUI actions to system behaviors, or translating users' natural-language utterances into GUI actions (e.g., typing ``Yes'' is equivalent to pressing the ``Yes'' button). This mapping overlooks the potential to adapt GUIs or streamline workflows based on contextual information gathered during the conversation~\cite{chen2025generative,luger2016having}. For example, a user who told a CAG they were purchasing two tickets for a date night could have those tickets pre-selected during the ticket selection stage, based on the previous conversation. As a further example, consider a CAG that supports troubleshooting a Wi-Fi connectivity problem: a user who says they restarted their router may skip a binary Yes/No button press asking whether they did so, which is typically the first question of the fixed GUI workflow. This potential is particularly pronounced in preference-driven, multi-step tasks like booking a movie ticket: earlier choices (theater, date) constrain later options (IMAX availability), often forcing users to backtrack~\cite{elasri2017frames,bursztyn2021developing}. 
The CAG can provide agentic decision support~\cite{demberg2011strategy}, proactively reflecting user preferences in the current GUI and steering subsequent workflow steps.

To address these gaps, we present \textbf{\sys{}} (\textbf{M}ultimodal \textbf{A}gent \textbf{E}mpowering \textbf{S}election by \textbf{T}ailoring GUIs, \textbf{R}ecalling preferences, and \textbf{O}rchestrating exploration), which supports users' decision making through GUI adaptation and workflow navigation assistance. 
\sys{} maintains a \textit{Preference Memory} that structurally extracts preferences from users' natural-language utterances and provides two capabilities grounded in this memory. \textbf{Preference-Grounded GUI Adaptation} applies in-place operators (augment, sort, filter, highlight) to manipulate information representation within the existing GUI without altering its structure, so that users can be informed about their available choices within the context of presented GUIs, making it easy to select and compare options. \textbf{Preference-Guided Workflow Navigation} detects conflicts between preferences and available options, proposes a specific step to return to, and logs the current path to avoid revisiting the failed paths. \rev{While recent systems externalize user intent into structured representations that generate new interfaces~\cite{cao2025generative,lam2026justintime}, \sys{} uses its Preference Memory to adapt an existing GUI in place, preserving the familiar context of a structured workflow.}
% In both conditions of our study, the agent can observe not only the options displayed on screen but also hidden attributes such as running times and ratings; the difference lies in how this information is delivered---as text advice in the Baseline or as in-place GUI modifications in \sys{}.

We investigate two research questions:
\begin{itemize}[leftmargin=1.5em]
\item \textbf{RQ1} How do preference-based GUI adaptation and navigation guidance in CAGs affect decision-making performance?
\item \textbf{RQ2} How do users perceive and experience such agentic interventions by CAGs?
% \item \textbf{RQ3} How does interaction modality (text vs.\ speech) influence decision-making performance (RQ1) and users' perceptions of the agent (RQ2)?
% \begin{itemize}
%   \item \textbf{RQ1} (Preference-Grounded GUI Adaptation):
%     What difference does in-place GUI adaptation make, compared to text advice, in helping users compare options and identify trade-offs within a stage?
%   \item \textbf{RQ2} (Preference-Guided Workflow Navigation):
%     What difference does structured preference extraction and workflow navigation make in users' backtracking, alternative exploration, and final decision outcomes?
\end{itemize}
\noindent While it is not our central research question, we also investigate how interaction modality (text vs.\ speech) influences decision-making performance and users' perceptions of the agent as a subquestion of RQ1 and RQ2.

To answer these research questions, we evaluated \sys{} through a $2\times2$ within-subjects study ($N=33$) using a movie-ticketing CAG, crossing Condition (Baseline vs. \sys{}) with Mode (Text vs.\ Voice). \rev{Mode is a secondary factor: prior work shows that interacting with an agent by voice rather than through a screen changes how actively users engage with agent-assisted tasks~\cite{reicherts2022good}, and speech may amplify the effect of \sys{} by making preferences easier to express, particularly in hands-free contexts such as automotive UI, smart displays, or AR glasses.} In the Baseline condition, the agent delivers information as textual advice within a single chat-stream layout with GUIs and without Preference Memory. In the \sys{} condition, the agent additionally adapts a persistent GUI panel through in-place operators and steers exploration via Preference Memory. 
% Each participant completed one warm-up task to become familiar with the target condition and one main task for evaluation, which required revisiting previous steps due to preference conflicts.
Each task involved a set of requirements that a participant needed to satisfy (e.g., a kid-friendly movie on Saturday for three at the closest theater). We evaluated performance through task success, preference violations, unpreferred selection count, task completion time, utterance pattern analysis, and subjective measures, including perceived usefulness, perceived burden, and preference rankings.
% The overarching research questions are as follows.
% \vspace{-5pt}
% Our study reveals three key findings. First, \sys{}'s GUI adaptation shifted how users communicated with the agent: information-seeking utterances decreased significantly ($p<.001$) while the absolute number of preference statements rose, and unpreferred selections were reduced ($p=.010$). Deliberation time increased under \sys{} ($p<.001$), indicating more engaged comparison rather than faster decisions. Second, workflow navigation was perceived as helpful ($M=5.81/7$, $p<.001$), though objective exploration paths did not differ significantly between conditions; a post-hoc analysis revealed that preferences expressed through action rather than language limited the system's backtrack accuracy. Third, voice modality elicited $40\%$ more utterances with substantially richer preference language ($p<.001$), but the Voice~+~\sys{} combination incurred the highest perceived burden across multiple dimensions (time, mental effort, learning difficulty, information overload). Despite comparable per-trial usefulness ratings, participants clearly preferred \sys{} over Baseline in both Text ($p<.001$) and Voice ($p=.019$) modes when asked to rank all conditions retrospectively. \sang{this needs update.}

Our study yields three key insights. First, supporting preference-aware interaction through GUI adaptation \rev{yields measurable reductions in preference violations, both in the final outcome and during the decision process}. Second, while it did not accelerate task completion, the system promoted more engaged decision-making, shifting user behavior toward expressing preferences and actively navigating the workflow. Third, interaction modality plays a critical role: while voice facilitates preference expression to a greater extent, it also introduces additional burden due to delays and turn-taking constraints. These findings demonstrate that integrating preference-aware adaptation and workflow guidance can be a new avenue for agentic behaviors of conversational agents with GUIs, highlighting both its benefits and design trade-offs.
The research contributions of this paper include:

% \vspace{-5pt}
\begin{enumerate}[leftmargin=2em]
  \item Design and development of \sys{}, a novel conversational agent that facilitates decision making in CAGs, equipped with GUI adaptation operators (augment, sort, filter, highlight) and preference-guided workflow navigation.
  % , with distinct adaptation policies for hard and soft preferences (\textbf{Preference-Grounded GUI Adaptation}).
  % \item A module that structurally extracts user preferences, detects cross-stage conflicts, and controls subsequent navigation steps (\textbf{Preference-Guided Workflow Navigation}).
  \item Empirical findings from a controlled within-subjects study that evaluates the effects of agentic decision support in CAGs.
\end{enumerate}

%% ============================================================
\section{Related Work}
\label{sec:related-work}
%% ============================================================

%% ------------------------------------------------------------
\subsection{Agents in GUI-Based Conversational Systems}
\label{sec:rw-agent-gui}
%% ------------------------------------------------------------

% The most prevalent paradigm casts the agent as an autonomous operator that interprets natural-language instructions and directly manipulates a graphical interface on the user's behalf. 
Agents have been used as autonomous operators that manipulate GUI on natural language input on the user's behalf~\cite{zhang2025largea,wang2025large}.
% This paradigm has expanded rapidly, 
Yet these agents share some limitations: they do not involve users at important decision points, and hallucinations remain a significant concern on complex, multi-step tasks~\cite{zou2025survey,zhang2025characterizing}.
% susceptibility to 
To restore user control, some systems introduce pause-and-override mechanisms. MIWA~\cite{chen2023miwa} supports step-through debugging and refinement in web automation, CowPilot~\cite{huq2025cowpilot} lets users pause or override agent actions during web navigation, and Morae~\cite{peng2025morae}  pauses execution at detected decision points. 
While these systems help support user control, they remain automation-centric: agent execution is the default mode, and user intervention occurs only at identified breakpoints rather than in a collaboration-centric manner.

A parallel stream grounds conversational interaction in the GUI itself. Weidele et al.~\cite{weidele2024empirical} study conversational control of GUIs in semantic automation, META-GUI~\cite{sun2022metagui} links GUI elements with dialogue context, and MALACHITE~\cite{ruoff2025malachite} provides a GUI-aware natural-language interface for complex applications. 
Traditional task-oriented dialogue systems similarly rely on intent detection and slot filling to guide task completion~\cite{louvan2020recent, weld2022survey}. 
Although these systems connect language to interface elements, the agent still mainly executes user intent within a fixed GUI. 
Recent work has also used LLMs to generate interface components or whole interfaces dynamically~\cite{hojo2025generativegui, chen2025generative, cao2025generative, amin2025promptcanvas, nandy2024bespokea}.
\rev{Such generation is increasingly guided by structured representations of user intent, including evolving task-driven data models~\cite{cao2025generative} and just-in-time objectives~\cite{lam2026justintime}, which define the task at hand and seed the synthesis of a new interface.}
However, generating \emph{new} interfaces rather than adjusting an existing GUI in place disconnects the user from the familiar context of the workflow.
% , which is an important drawback when the task involves a service with an established interaction flow.

A growing body of work instead adapts \emph{existing} interfaces through natural language rather than generating new ones at each turn. 
Stylette~\cite{kim2022stylettea} maps styling goals to CSS edits, DynaVis~\cite{vaithilingam2024dynavisa}  creates manipulable widgets for visualization editing, and DirectGPT~\cite{masson2024directgpt} supports in-place modification of selected objects. These systems show that natural language can support in-situ GUI changes, but each interaction is largely self-contained. 
Recent works such as IRF~\cite{peng2025morae}  and CARE~\cite{peng2025navigating} explored sustained interaction by updating interface content as users refine preferences over time. Still, these systems largely position the agent as the primary executor of user intent.

% A growing body of work takes a middle path, adapting  interfaces through natural-language input rather than generating new ones wholesale. Several systems translate individual natural-language utterances into concrete visual modifications: Stylette~\cite{kim2022stylettea} maps high-level styling goals to CSS edits on live web pages, DynaVis~\cite{vaithilingam2024dynavisa} synthesizes persistent, manipulable widgets from editing requests for data visualizations, and DirectGPT~\cite{masson2024directgpt} lets users select on-screen objects and issue short prompts to modify them in place. These tools demonstrate that natural language can serve as an effective channel for in-situ GUI modification, yet each interaction is self-contained: the system responds to one command without carrying forward context from earlier exchanges. A step closer to sustained interaction, IRF~\cite{tang2025interactive} embeds natural-language commands within a conventional recommendation feed and uses a dual-agent architecture to parse preferences and adjust ranking policies on the fly, while CARE~\cite{peng2025navigating} pairs a chat panel with structured solution and needs panels that co-evolve as users iteratively refine exploratory queries. Taken together, these systems integrate language into GUI interaction in different ways, but they continue to center the agent as the primary executor of user intent.

%% ------------------------------------------------------------
\subsection{Decision Support with Preference Management}
\label{sec:rw-preference-decision}
%% ------------------------------------------------------------
Conversational recommender systems (CRSs) have long studied how to elicit and refine user preferences through dialogue~\cite{thompson2004personalized,christakopoulou2016conversational,jannach2022survey}. Recent LLM-based systems extend this work with stronger preference elicitation, explanation, and recommendation~\cite{feng2023large,gao2023chatrec,kook2025empowering}. 
For example, RecLLM~\cite{friedman2023leveraging} builds user profiles from conversation history to personalize recommendations. 
However, these systems mainly operate within a single recommendation stage, where one round of
elicitation leads to one set of results.  
In multi-step tasks such as travel booking, earlier selections constrain later options, and users must compare alternatives, revisit previous choices, and balance competing preferences across stages~\cite{elasri2017frames,bursztyn2021developing}. 
\citeauthor{qin2025compass} framed this complexity as a constrained optimization problem and found that LLMs are good at following hard constraints, but perform poorly in following soft preferences~\cite{qin2025compass}.
This motivates further research in designing CUIs for preference-driven multi-step tasks.

% Conversational recommender systems (CRSs) have long explored how to elicit and refine user preferences through dialogue~\cite{thompson2004personalized,christakopoulou2016conversational,jannach2022survey}. Building on this line of work, recent LLM-based systems have substantially expanded capabilities in preference elicitation, explanation, and recommendation~\cite{feng2023large,gao2023chatrec,kook2025empowering}. For example, RecLLM~\cite{friedman2023leveraging} constructs interpretable natural-language user profiles from conversation history and feeds them to an LLM to personalize session-level recommendations. These systems, however, operate within a single recommendation stage, where one round of elicitation leads to one set of results. In multi-step tasks such as travel booking, earlier selections constrain later options, and users must compare alternatives, revisit previous choices, and balance competing preferences across stages~\cite{elasri2017frames,bursztyn2021developing}. COMPASS~\cite{qin2025compass} formalizes this setting as a constrained optimization problem with hard constraints and soft preferences, and finds that LLM agents reliably satisfy hard constraints but fail to optimize soft preferences, highlighting that preference-driven multi-step tasks remain an open challenge.

For cross-stage preference management, the system must track preferences as they change over longer conversations.
Recent evaluations suggest this remains difficult. PrefEval~\cite{zhao2025llms} shows that preference-following accuracy drops below 10\% by 10 turns in zero-shot settings.
PERSONAMEM~\cite{jiang2025know} finds that frontier models reach only around 50\% accuracy in recognizing dynamic profile changes, and CUPID~\cite{kim2025cupid} reports under 50\% precision in inferring contextual preferences from multi-turn histories.
These results motivate structured external representations instead of relying only on the LLM context window. 
Recent works have used natural-language records, reflection-driven summaries, and confidence-weighted propositions about user behavior~\cite{park2023generativea, shaikh2025creatinga}. 
These systems show that external representations support more reliable capture and retrieval of user information, guiding the way for dedicated preference stores for sustained,
preference-aware interaction in multi-step settings.

% For such cross-stage preference management to work, the system must reliably track preferences as they evolve over extended conversations, an ability that recent evaluations call into question. PrefEval~\cite{zhao2025llms} shows that preference-following accuracy drops below 10\% at merely 10 conversation turns in zero-shot settings. PERSONAMEM~\cite{jiang2025know} reveals that frontier models achieve only around 50\% accuracy in recognizing dynamic changes in user profiles over time, and CUPID~\cite{kim2025cupid} finds that precision in inferring contextual preferences from multi-turn interaction histories falls under 50\%. These findings motivate structured external representations rather than relying on the LLM's context window alone. Research on agent memory offers one such path~\cite{zhang2025survey}: Generative Agents~\cite{park2023generativea} store experiences as natural-language records and generate higher-level summaries through reflection, and the General User Model (GUM)~\cite{shaikh2025creatinga} constructs confidence-weighted propositions about user behavior and preferences from unstructured observations. These systems demonstrate that structured external representations can capture and retrieve user information more reliably than in-context tracking, motivating dedicated preference stores as a necessary foundation for sustained, preference-aware interaction in multi-step settings.

The remaining challenge is how to apply stored preferences during the task, both for interface \emph{adaptation} and workflow \emph{navigation}. 
Within a stage, option presentation affects decision quality: structuring choices around a preference model and showing trade-offs improves spoken-dialog performance~\cite{demberg2011strategy,dutton2001amount}, while in visual interfaces, format, sorting, and filtering affect search time and decision confidence~\cite{hong2004designing,thai2012visual,wan2009paradoxical}.
Nav Nudge~\cite{yu2024reducing} combines voice input with an LLM to highlight relevant options on a mobile GUI.
Across stages, accumulated preferences may conflict with available options, forcing the system to decide how to proceed. 
The Frames corpus~\cite{elasri2017frames} shows that users naturally revisit prior options and compare alternatives, making frame tracking a core dialogue capability.
However, these adaptations are driven by fixed rules or direct manipulation rather than by an agent interpreting ongoing dialogue.

% The remaining challenge is how to apply stored preferences during the task, both to adapt the interface and to navigate the workflow. 
% Within a given stage, how options are presented significantly affects decision quality: structuring options based on a preference model and explicitly presenting trade-offs improves task success in spoken dialog~\cite{demberg2011strategy,dutton2001amount}, while in visual interfaces, format, sorting, and filtering shape search time and decision confidence~\cite{hong2004designing,thai2012visual,wan2009paradoxical}. Nav Nudge~\cite{yu2024reducing} combines voice input with an LLM to visually reduce the search space by highlighting relevant options on a mobile GUI. Yet these adaptations are driven by fixed rules or direct manipulation rather than by an agent interpreting ongoing dialogue. Across stages, accumulated preferences may conflict with available options, requiring the system to decide how to proceed. The Frames corpus~\cite{elasri2017frames} reveals that users naturally switch between previously discussed options and compare attributes across alternatives, establishing frame tracking as a core dialogue capability. DGDVA~\cite{tiwari2021dynamic} detects goal discrepancies via user sentiment and dynamically updates the agent's objective, and AWARE-US~\cite{kurmaz2026awareus} treats infeasibility as preference-aware query repair, relaxing the least important constraints. However, these conflict-resolution mechanisms operate within a single domain or query.

%% ------------------------------------------------------------
\subsection{Summary}
\label{sec:rw-summary}
%% ------------------------------------------------------------

Across the approaches reviewed in \autoref{sec:rw-agent-gui} (autonomous execution, mixed-initiative pausing, GUI-aware dialogue, generative interfaces, and language-driven adaptation), agents either act on behalf of the user within an existing GUI, generate entirely new interfaces, or adapt a single interface state in response to one-shot commands; none adaptively modify an existing GUI based on preferences accumulated over a multi-turn conversation to support ongoing decision making. \rev{Unlike intent representations that define a task and seed the generation of its interface~\cite{cao2025generative, lam2026justintime}, \sys{}'s Preference Memory augments and curates an existing GUI and steers navigation within its fixed workflow.} Research on preference management (\autoref{sec:rw-preference-decision}) provides elicitation loops, memory mechanisms, and local conflict-handling strategies, but these remain confined to single-stage settings or isolated queries, and LLMs alone cannot reliably sustain preference tracking over extended conversations. \sys{} bridges these two streams: it interprets preferences expressed in dialogue to apply in-place GUI adaptation operators (augment, sort, filter, highlight), and it structurally tracks preferences across workflow stages to detect conflicts, propose backtracking, and remember failed paths.

%% ============================================================
\section{System Description: \sys{}}
\label{sec:maestro}
%% ============================================================

\sys{} comprises three modules built upon a Conversational Agent with GUI (CAG):
% for movie ticketing (\autoref{sec:target-domain}):
an agent architecture that serves as the base system; a Preference Memory that extracts user preferences as structured records and maintains them throughout the session (\autoref{sec:pref-memory}); and two modules informed by Preference Memory, Preference-Grounded GUI Adaptation (\autoref{sec:gui-adaptation}) and Preference-Guided Workflow Navigation (\autoref{sec:exploration-steering}). The agent architecture alone constitutes a fully functional CAG; \sys{} extends it with the preference memory and both modules.
All inference tasks described in this section, including generating responses, extracting preferences from conversations, deciding to adapt GUIs, and guiding navigation steps, are powered by OpenAI GPT-5.4.
\subsection{Target Domain: Movie Ticketing Assistant}
\label{sec:target-domain}
%% ------------------------------------------------------------

While the approach is applicable to domains where a CAG can be used, we focus on a specific domain—a movie ticketing agent—to demonstrate it. We introduce the target domain here before describing the system, as the examples will be used in the system description.

The system follows a linear multi-stage workflow (movie, theater, date, time, seat, and confirmation), where a user has to select one option at each stage \rev{(at the seat stage, one or more seats)}. Each stage is associated with a dedicated GUI component: button groups for movie, theater, and time selection; a calendar for date selection; and a seat map for choosing seats. This workflow is modeled after the information structure of major movie-ticketing services such as Fandango or AMC; stages that co-occur on a single screen in production services (e.g., theater, date, and showtime) are separated into individual stages so that each decision point can be addressed independently. Users may navigate backward to revise earlier choices, and skipping or reordering stages is not permitted.

\begin{figure}[t]
  \centering
  \includegraphics[width=0.92\columnwidth, trim=0 8 0 16, clip]{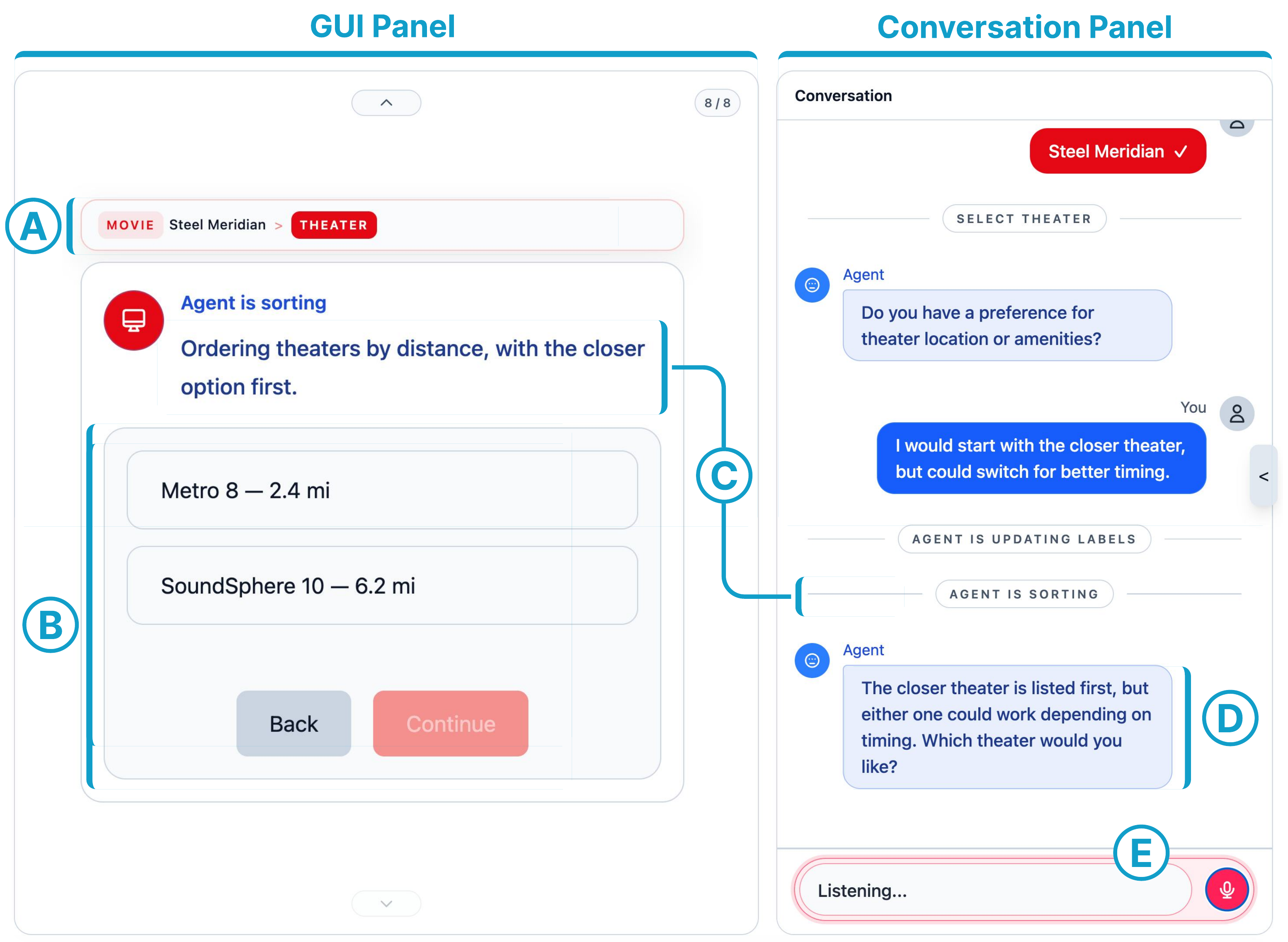}
  \vspace{-6pt}
  \caption{\sys{} layout at the theater selection stage. 
  % A split-panel layout with a GUI panel (left) and a conversation panel (right).
  \textbf{\circnumblue{A}}~Workflow breadcrumb showing current progress. 
  % \textbf{\circnumblue{A}}~A workflow breadcrumb displaying the current stage and selections made so far (e.g., Movie: Steel Meridian $>$ Theater). 
  % \textbf{\circnumblue{B}}~The GUI panel displays the current stage's options after adaptation; here, distances have been augmented and options sorted by proximity, with a descriptive overlay message (``Agent is sorting''). 
  \textbf{\circnumblue{B}}~The GUI shows distance augmentation and proximity-based sorting. 
  \textbf{\circnumblue{C}}~Adaptation actions shown in the chat panel. 
  \textbf{\circnumblue{D}}~Agent follow-up summarizing its actions and information, prompting the next decision. 
   % \textbf{\circnumblue{C}}~Each adaptation action is communicated through a paired representation: a descriptive overlay in the GUI panel linked to a corresponding badge in the conversation panel. 
  % \textbf{\circnumblue{D}}~The agent's follow-up message in the conversation panel, synthesizing the adaptation result and prompting the user's next decision. 
  \textbf{\circnumblue{E}}~Voice input interface.}
  \Description{A split-panel layout for the MAESTRO condition at the theater selection stage. The left GUI panel shows the result after two adaptation actions (augment and sort) have been applied. A workflow breadcrumb at the top reads Movie: Steel Meridian, Theater. Below it, a red overlay message reads ``Agent is sorting --- Ordering theaters by distance, with the closer option first.'' Two theater buttons appear in sorted order: Metro 8 --- 2.4 mi first, then SoundSphere 10 --- 6.2 mi, followed by Back and Continue buttons. The right conversation panel shows a chronological flow: a completed Steel Meridian badge at the top, a Select Theater stage label, the agent asking ``Do you have a preference for theater location or amenities?'', the user's reply ``I would start with the closer theater, but could switch for better timing.'', two adaptation badges (Agent is updating labels and Agent is sorting) connected to the GUI panel overlay by a visual link line, and the agent's follow-up response ``The closer theater is listed first, but either one could work depending on timing. Which theater would you like?'' A voice input bar with ``Listening...'' and a microphone button appears at the bottom.}
  \label{fig:maestro-layout}
  \vspace{-10pt}
\end{figure}

\subsection{Interface Layout}

\sys{} divides the screen into a GUI panel on the left and a conversation panel on the right (\autoref{fig:maestro-layout}). The GUI panel persistently displays the current stage’s options and the agent’s adaptation actions with smooth animation. (see~\autoref{sec:appendix-screenshot})
While typical CAG designs present a linear stream of chat and GUI elements within a scrollable window, we found that this approach can make it difficult for users to comprehend the interface. In particular, GUI elements may be pushed out of view as users and the agent add new chat messages, or the GUI adaptation module may append entirely new GUI containers, making the intermixed history of GUI and chat unnecessarily lengthy. Placing the GUI panel on one side ensures users can immediately see what has changed.
Scrolling the conversation panel synchronizes the GUI panel, allowing users to review previous GUI states associated with each chat message. This preserves the ability to inspect conversation history that users expect from typical chat interfaces, where prior messages and associated GUI states remain accessible. 
% The specific mechanisms by which adaptation actions are communicated across the two panels are described in \autoref{sec:gui-adaptation}.

% NOTE: "Controlled Evaluation Environment" content moved to Study section
% Original: This domain is a simplified abstraction of web and mobile multi-step interaction patterns.
% In actual web environments, additional cognitive load arises from page transitions and the effort
% of locating required UI elements; these factors are excluded here. The environment is designed as
% a controlled setting in which the effects of adaptation and workflow navigation can be evaluated
% independently of such confounds.

%% ------------------------------------------------------------
%% ------------------------------------------------------------
\subsection{Preference Memory}
\label{sec:pref-memory}
%% ------------------------------------------------------------

Preference Memory serves as the shared state for Preference-Grounded GUI Adaptation (\autoref{sec:gui-adaptation}) and Preference-Guided Workflow Navigation (\autoref{sec:exploration-steering}). Preferences are extracted from users' natural-language utterances as structured records and maintained throughout the session, so that the current screen's representation and the navigation path are consistently grounded in the same preference state.

%enabling both modules to reference the same preferences and consistently control the current screen's representation and the navigation path.

\paragraph{Extraction and update.}
Preference extraction is invoked on every user message. The underlying system prompt of the agent is designed to elicit users' preferences (e.g., ``Do you have a preference for theater location or amenities?", See~\autoref{fig:maestro-layout}).
%Pilot observations revealed a tendency for users to proceed without explicitly stating their preferences. To compensate, the agent proactively asks about preferences to increase the probability that a user shares what they need. (e.g., ``I prefer a multiplex with free parking.'') 
While merely selecting an option is not recognized as a preference, explicit expressions of intent, such as constraints and desires, are stored as preferences.
When a new preference overlaps or conflicts with an existing one, the record's preference is updated.
%The LLM prompt used to extract preferences is included in the supplementary material and returns the following object in JSON format. 
% GUI clicks are not treated as preferences because a click may represent an exploration action rather than a solid intent. 

\paragraph{Preference Record Structure}
Each preference is represented as a record with three properties. The \texttt{description} field captures the preference as a natural-language description. The \texttt{strength} field is either \texttt{hard} (must be satisfied; cued by words such as ``must,'' ``only,'' ``need'') or \texttt{soft} (satisfy if possible; cued by ``prefer,'' ``ideally,'' ``want''); ambiguous cases default to soft. The strength property is later used to determine which adaptation policy to use. The \texttt{relevantStages} field lists the workflow stages to which the preference applies, assigned by an LLM agent. 
% using the per-stage attribute guide described in \autoref{sec:agent}. 
An example, included in \autoref{sec:JSON}, illustrates a representative set of preference records.

% 507f1f77bcf86cd799439011
% 65e9b2c4a3d1f8e7b6c5a2d3

% When a user changes an existing preference, the record is updated in place. For example, a soft preference restated without hedging words (e.g., ``I prefer comedies'' later rephrased as ``It has to be a comedy'') is upgraded to hard, and withdrawn preferences are removed.

%% ------------------------------------------------------------
\subsection{Preference-Grounded GUI Adaptation}
\label{sec:gui-adaptation}
%% ------------------------------------------------------------

Preference-Grounded GUI Adaptation dynamically changes the presentation of GUI elements based on the preferences stored in Preference Memory.
%This subsection describes the four types of adaptation operators, the policy that selects among them, and how adaptations are communicated to the user.

\begin{figure*}[t!]
  \centering
  \includegraphics[width=\textwidth]{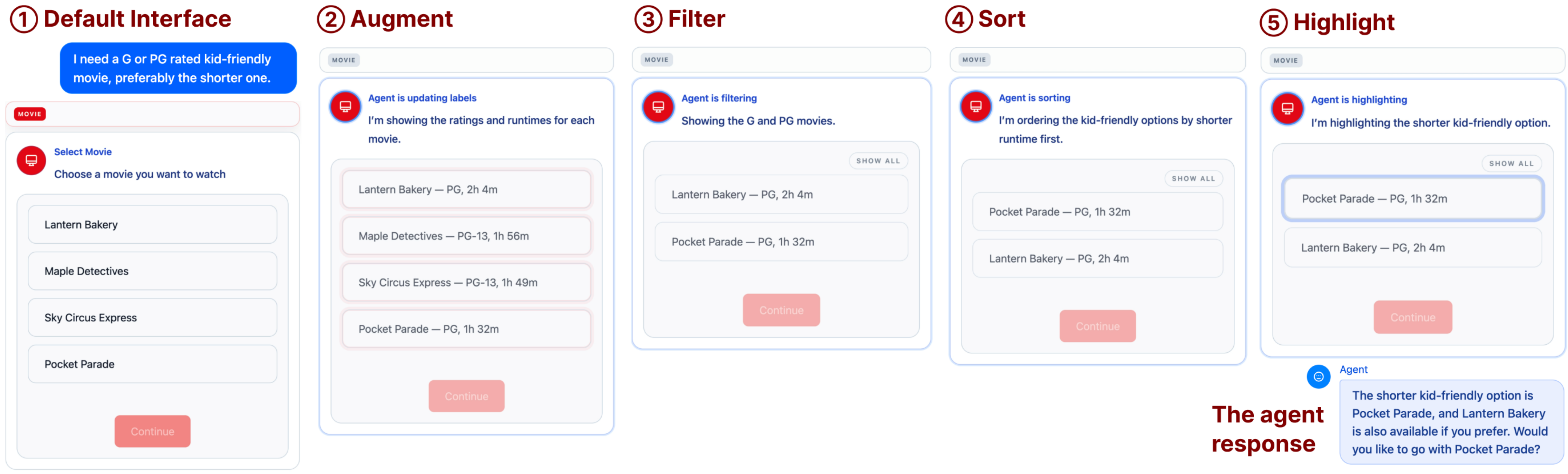}
   \caption{
  % The four-step adaptation policy applied to a movie-selection example.
  Four-step adaptation policy illustrated on a movie-selection example. 
  % \textbf{\circnum{1} Default Interface:} The user states two preferences: ``G or PG rated kid-friendly movie'' (\texttt{hard}) and ``preferably the shorter one'' (\texttt{soft}), with the default GUI displaying four movie buttons with titles only. 
  \textbf{\circnum{1} Default interface}
  \textbf{\circnum{2} Augment:} Ratings and runtimes  appended to each button. 
  \textbf{\circnum{3} Filter:} PG movies  filtered out 
  % \textbf{\circnum{4} Sort:} The remaining movies are sorted by runtime in ascending order.
    % \textbf{\circnum{5} Highlight:} The shortest option (Pocket Parade) is highlighted, and the agent poses a confirmation question asking whether the user would like to go with Pocket Parade.
  \textbf{\circnum{4} Sort:} Filtered movies are sorted by runtime.
  \textbf{\circnum{5} Highlight:} Shortest option highlighted with agent confirmation.}

  \Description{A five-panel walkthrough of the four-step adaptation policy at the movie selection stage, followed by an agent response. Panel 1, labeled Default Interface, shows a user speech bubble reading ``I need a G or PG-rated kid-friendly movie, preferably the shorter one.'' Below, the GUI displays a ``Select Movie'' header with the instruction ``Choose a movie you want to watch'' and four movie buttons with titles only: Lantern Bakery, Maple Detectives, Sky Circus Express, and Pocket Parade, with a Continue button at the bottom. Panel 2, labeled Augment, shows an ``Agent is updating labels'' overlay with the message ``I'm showing the ratings and runtimes for each movie.'' The four buttons now display appended metadata: Lantern Bakery --- PG, 2h 4m; Maple Detectives --- PG-13, 1h 58m; Sky Circus Express --- PG-13, 1h 49m; Pocket Parade --- PG, 1h 32m. Panel 3, labeled Filter, shows an ``Agent is filtering'' overlay with the message ``Showing the G and PG movies.'' Only two buttons remain after PG-13 movies are removed: Lantern Bakery --- PG, 2h 4m, and Pocket Parade --- PG, 1h 32m, with a Show All link above. Panel 4, labeled Sort, shows an ``Agent is sorting'' overlay with the message ``I'm ordering the kid-friendly options by shorter runtime first.'' The two movies are reordered: Pocket Parade --- PG, 1h 32m first, then Lantern Bakery --- PG, 2h 4m, with a Show All link. Panel 5, labeled Highlight, shows an ``Agent is highlighting'' overlay with the message ``I'm highlighting the shorter kid-friendly option.'' Pocket Parade is visually emphasized with a colored border, with a Show All link. Below the five panels, an agent response chat bubble reads ``The shorter kid-friendly option is Pocket Parade, and Lantern Bakery is also available if you prefer. Would you like to go with Pocket Parade?''}
  \label{fig:gui-adaptation}
  \vspace{-6pt}
\end{figure*}

\subsubsection{Adaptation Operators}

To support decision-making, \sys{} reorganizes the presentation of information based on the current GUI structure and the user's preferences so far. Instead of generating GUI elements from scratch, we developed four GUI adaptation patterns that update GUI elements in place. 
%The following four patterns were derived from information presentation and adaptive UI research~\cite{demberg2011strategy,hong2004designing,thai2012visual,yu2024reducing}.
% \sang{why is this related to preference? This section can be presented "before" preference extraction/recording.}

\emph{Augment} operator appends additional attributes to the text displayed on each button. As described in \autoref{sec:target-domain}, each button shows only a single key attribute by default; the Augment operator augments text with metadata available for an item (e.g., audience ratings next to movie titles, or screening end times alongside showtimes). Surfacing all decision-relevant attributes at once, rather than requiring users to request them incrementally, improves task completion and satisfaction~\cite{dutton2001amount}.
%The Augment operator is always accompanied by the other three adaptation operators, as the metadata provides reasoning behind the adaptation. For example, filtering out R-rated movies can be communicated by displaying each movie's rating next to its title. See~\autoref{fig:gui-adaptation}.

The \emph{Filter} operator curates information by filtering out items that do not meet a user's requirements. For example, it hides movies that are not suitable for all ages when a user specifies their preference for a family movie night with a 6-year-old kid. Filtering is commonly used and validated in dynamic query and faceted search research~\cite{thai2012visual}. 
%It hides elements that do not meet a specified condition, displaying only the qualifying subset; 

\emph{Sort} operator reorders the same set of items according to a selected attribute, especially when the selected attributes are numeric. An example is reordering the movie list by audience rating when a user asks for ratings of the movies on screen. According to the literature, sorting facilitates comparative judgment~\cite{hong2004designing}. 

\emph{Highlight} assigns visual emphasis by adding a colored border to one or more elements, directing attention without adding or removing information. Visually emphasizing relevant options has been shown to help users locate target features more quickly~\cite{yu2024reducing,yu2023where}. Examples include highlighting a recommended showtime or four adjacent seats in a seatmap.

These four adaptation operators span the space of in-place information transformations that preserve GUI structure (layout, components, and navigation remain intact). All four are implemented as lightweight frontend in-place modifications. 
%A declarative UI specification represents the interface as a structured description of components bound to an underlying data model, separating application state from its visual presentation, following modern front-end frameworks that architecturally divide application state (a JSON data object) from UI structure (declarative component definitions bound to that state). Thanks to the declarative UI specification with separate data and display layers, the agent can expose hidden attributes by augmenting and adjusting display order, visibility, and emphasis using the adaptation operators.

\subsubsection{Applying GUI Adaptation}

In this section, we explain how \sys{} applies GUI adaptation operators through an example shown in \autoref{fig:gui-adaptation}.  
% At each message that a user enters or the system transitions to a new stage of the workflow, \sys{} will detect a user's preference from the chat history. In the example shown in \autoref{fig:gui-adaptation}, the user states two preferences: G or PG rated kid-friendly movie'' (\texttt{hard}) and preferably the shorter one'' (\texttt{soft}). Then 

% \paragraph{Step~1: Annotating with meta-data.}
At each stage, \sys{} determines which adaptation operators to apply to the current GUI elements based on the relevant preferences logged in Preference Memory.
In~\autoref{fig:gui-adaptation}-\circnum{1}, the user states two preferences: ``G or PG rated kid-friendly movie'' (\texttt{hard}) and ``preferably the shorter one'' (\texttt{soft}).
Once such preferences are identified, \sys{} will first \textit{augment} the GUI elements with metadata relevant to the preferences; each movie option will now have ratings and runtimes appended (\autoref{fig:gui-adaptation}-\circnum{2}).

Because the first preference's strength is hard, \sys{} applies the \textit{filter} operator to hide options that do not satisfy it, such as R-rated movies or showtimes that start after 5 PM. At stages where filtering is not applicable (Calendar and Seat-map stages), highlight is used instead to mark all matching items. The user can always restore filtered-out items using the ``Show All'' button at the top-right of the UI container. (\circnum{3}).

For the \texttt{soft} preference, which reflects a user's preference for shorter movies, \sys{} applies the \textit{Sort} operator and uses metadata to reorder items (\circnum{4}). The \textit{Sort} operator is effective when the metadata type is ordinal or numeric, such as ratings, price, and distance. The optimal option is then highlighted (\circnum{5}).
Categorical metadata, such as genre or screen type (e.g., IMAX), cannot be used with the Sort operator; instead, the \textit{Highlight} operator visually emphasizes options that satisfy a user's soft preference.
%Each adaptation action is rendered with real-time animation (e.g., items switching positions slowly for Sort operator) in the GUI panel and communicated to the conversation panel as a badge indicating the action type (\autoref{fig:maestro-layout}-\circnumblue{C}). 

Lastly, once GUI adaptation is complete, the \sys{} agent will pose a confirmation question in a follow-up message (e.g., ``Would you like to go with this option?''). Rather than silently waiting for a selection, the agent proactively proposes the outcome of its adaptation and invites the user to accept, reject, or revise it. Not all four operators apply in every scenario; however, when any of the other three operators is used, the Augment operator always accompanies it, surfacing the metadata that justifies the adaptation.

The Preference Memory enables reapplying the GUI adaptation. For example, when a user wants to watch a movie on an IMAX screen and later selects another movie, \sys{} filters out showtimes that are not on IMAX screens during the showtime stage. 

\subsection{Preference-Guided Workflow Navigation}
\label{sec:exploration-steering}
%% ------------------------------------------------------------
When \sys{} guides users through a workflow based on their preferences, conflicts can arise when those preferences (e.g., two adjacent seats) cannot be met due to real-world constraints (e.g., all available seats are single seats). In such cases, users must seek alternatives and choose different paths. However, they may forget which step to return to in order to find alternatives that satisfy other preferences, or attempt to revisit a dead end they have already encountered.
%Paths that lead to such conflicts are recorded to prevent them from being repeated in subsequent navigation. Without this mechanism, the agent may repeatedly guide users along the same path due to Preference-Grounded GUI Adaptation, only to encounter the same conflict.
To address this issue, the Preference-Guided Workflow Navigation module assesses the viability of the current navigation path and recommends a step the user should return to when conflicts arise.

% When relying solely on conversation history, the likelihood that the agent fails to reflect previously failed paths in its planning and actions increases as the conversation grows. Because the full context is provided to the language model at every turn, consistently tracking which paths have already been tried and determining how far back to backtrack becomes unreliable. 
% \sys{} maintains dead-end records and alternative counts as explicit structured objects that are injected into the agent's input, separate from the conversation history.

% I can't compile :( because of errors that this line is making Good  Sorry Korean is making the error I guess.
\subsubsection{Backtracking Suggestions based on Alternative Counts}
\sys{} tracks the number of remaining alternatives at each stage along the navigation path. When computing these counts, it leverages the results of Preference-Grounded GUI Adaptation. For example, if three theaters are available and filtering by a free-parking preference narrows the set to two, the one remaining theater (excluding the user's current selection) is stored as the alternative count for the theater stage. This calculation applies when the agent performs a filter action; for highlight actions, it applies only when the highlight was used to reduce choices rather than for visual comparison. For instance, at the date stage, only the highlighted dates are counted as viable alternatives.

These per-stage alternative counts are injected into the agent's input when a conflict is detected, so the agent can suggest an appropriate step to backtrack to. For example, if the agent perceives that four adjacent seats are unavailable at the current seat-selection stage, it receives the alternative counts from all preceding stages. If the user had indicated that only a specific date was acceptable (highlighting that date as the sole option), the date stage would have zero alternatives. If the theater stage preceding it had at least one alternative, the agent identifies the theater stage as the appropriate backtrack target and proposes returning to it, rather than going back to the date stage, which has no other alternatives. 

%Sang's version: For example, if no three adjacent seats are available, \sys{} determines whether alternative showtimes exist under the user’s preferences. If there was only one showtime that satisfies a user's preference on the dead-end path, the agent suggests returning two steps earlier (e.g., “Would you like to try a different date, a different theater, or a different movie?”) rather than proposing to return to the showtime schedule stage, which is already known to be unavailable. \sys{} takes this type of question as an opportunity to elicit a more concrete preference from a user. For example, if a user says ``It has to be this Saturday (\textit{hard preference}). Let's try another theater.", the agent will create a new preference (e.g., Date: Saturday), return to the theater selection stage, and gray out the theater that led to a dead-end path, facilitating their further navigation. 

\subsubsection{Dead-End Recording}
When the user accepts the backtrack suggestion due to the lack of a viable option, \sys{} records the current path as a dead end. This record is then injected into the agent's input on subsequent turns, so the agent can avoid re-exploring previously failed paths. Each dead end is scoped to the specific combination of preceding selections that produced it. For example, if selecting Movie A (Stage 1) and Theater B (Stage 2) yields no viable showtimes (Stage 3), Theater B is excluded on that path given Movie A, but it remains a valid option when a different movie is selected on Stage 1.

\subsubsection{Updating Alternative Counts and Dead-Ends When a Preference Changes}

Because alternative counts and dead-end records are each linked to a specific entry in Preference Memory, \sys{} can automatically update them when a preference changes. For example, if several paths were recorded as dead ends because no IMAX showtime was available, and the user later relaxes that constraint, the dead-end records tied to that preference (i.e., IMAX screen) are removed, and those paths become available for exploration again. Filter and highlight adaptations linked to the changed preference are likewise cleared, and alternative counts are recalculated accordingly. The agent then replans based on the updated alternative counts and dead-end records, without requiring the user to manually retrace their steps.

\subsection{Agent Architecture}
\label{sec:agent}

At each turn, \sys{} observes the current GUI state, including displayed options and any applied adaptations, together with underlying data attributes not visible in the interface (e.g., ratings, distance), the user’s utterance, and prior selections. Based on this, it selects actions via LLM tool calling from a context-dependent toolset that includes standard GUI interactions (selection, navigation) and the adaptation operators described in \autoref{sec:gui-adaptation}. Preferences expressed early are mapped to relevant stages via the \texttt{relevantStages} field in Preference Memory, so they can be evaluated when those stages are reached.

When a preference changes, related dead-end records and adaptations are automatically updated or removed, enabling the agent to replan without requiring the user to retrace steps. Because the Preference Memory carries the core decision context, the conversation history is \rev{truncated} to recent turns to mitigate the lost-in-the-middle problem~\cite{liu2024lost}, where language models miss intermediate context as conversations grow; the current stage’s GUI state is supplied separately as a representative snapshot.

\begin{revblock}
\subsection{Implementation}
\label{sec:implementation}

\sys{} consists of a React frontend that renders the GUI and conversation panels, a Fastify backend that serves the scenario database and session state, and an agent runtime connected to the frontend through a WebSocket relay. We implemented the agent's two LLM components, the preference extractor and the action planner, as structured prompts over OpenAI GPT-5.4. The extractor runs only in the \sys{} condition, and the planner's prompt is assembled from per-condition rule blocks. Each component calls the Responses API with temperature set to 0 and requests structured JSON output; the full system prompts appear in \autoref{sec:appendix-prompts}. Responses that fail schema parsing are salvaged when possible and otherwise discarded, leaving the interface state unchanged for that turn. Voice mode uses OpenAI gpt-4o-mini-transcribe for speech recognition and gpt-4o-mini-tts for spoken output. Our implementation is publicly available.\footnote{\url{https://github.com/wooogler/maestro-movie}}
\end{revblock}

\section{Evaluation Method}
\label{sec:evaluation-method}
%% ============================================================

To evaluate whether \sys{} improves users' decision-making compared to the current CAG paradigm, we conducted a within-subjects study in the movie-ticketing domain. 
%% ------------------------------------------------------------
\subsection{Study Design}
\label{sec:study-design}
%% ------------------------------------------------------------

\subsubsection{\texorpdfstring{\rev{$2 \times 2$ Within-Subjects Experimental Design}}{2 x 2 Within-Subjects Experimental Design}}

All participants experienced four combinations of two factors: \textbf{Condition} (Baseline vs.\ \sys{}) and \textbf{Mode} (Text vs.\ Voice). We denote each combination using the following labels: Baseline in Text mode (\textbf{BT}), Baseline in Voice mode (\textbf{BV}), \sys{} in Text mode (\textbf{MT}), and \sys{} in Voice mode (\textbf{MV}).

The primary comparison is Condition (Baseline vs.\ \sys{}), capturing the integrated effect of the agent's decision-support approach. The secondary comparison is Mode (Text vs.\ Voice), examining how input/output modality affects the decision process. The Condition~$\times$~Mode interaction is explored to determine whether Voice amplifies or modulates the effect of \sys{}. Because the same participant experiences all four cells, individual differences are controlled.

% All participants performed one warm-up task and one main task. Table~\ref{tab:conditions} summarises the two conditions.

% \begin{table*}[t]
% \caption{Summary of the two conditions. Both agents can query the same backend metadata; the difference is in how decision support is delivered.}
% \label{tab:conditions}
% \small
% \begin{tabular}{p{1.8cm} p{3.8cm} p{4.2cm} p{5.2cm}}
% \toprule
% \textbf{Condition} & \textbf{UI Layout} & \textbf{Preference Management} & \textbf{Agent Behaviour} \\
% \midrule
% Baseline &
%   Single stream---GUI elements and agent messages mixed in one chat timeline. &
%   Full conversation history + last GUI state only. Preferences exist only in chat history. &
%   Text advice, recommendations, and comparisons based on conversation history and backend metadata. No GUI modification. \\
% \addlinespace
% \sys{} &
%   Separated---persistent GUI panel (left) and chat panel (right). Adaptations reflected in real time. &
%   Preference Memory: structured JSON records (description, strength, relevantStages). Recent-turn condensation mitigates lost-in-the-middle. &
%   GUI Adaptation 
%   % (two-step policy: visibility gate $\to$ strength-based operator selection) 
%   + Workflow Navigation 
%   % (conflict detection, deferred dead-end recording, deterministic backtracking via alternative counts). 
%   \\
% \bottomrule
% \end{tabular}
% \end{table*}
\subsubsection{Condition: Baseline vs. \sys{}}
\begin{figure}[t]
  \centering
  \includegraphics[width=\columnwidth]{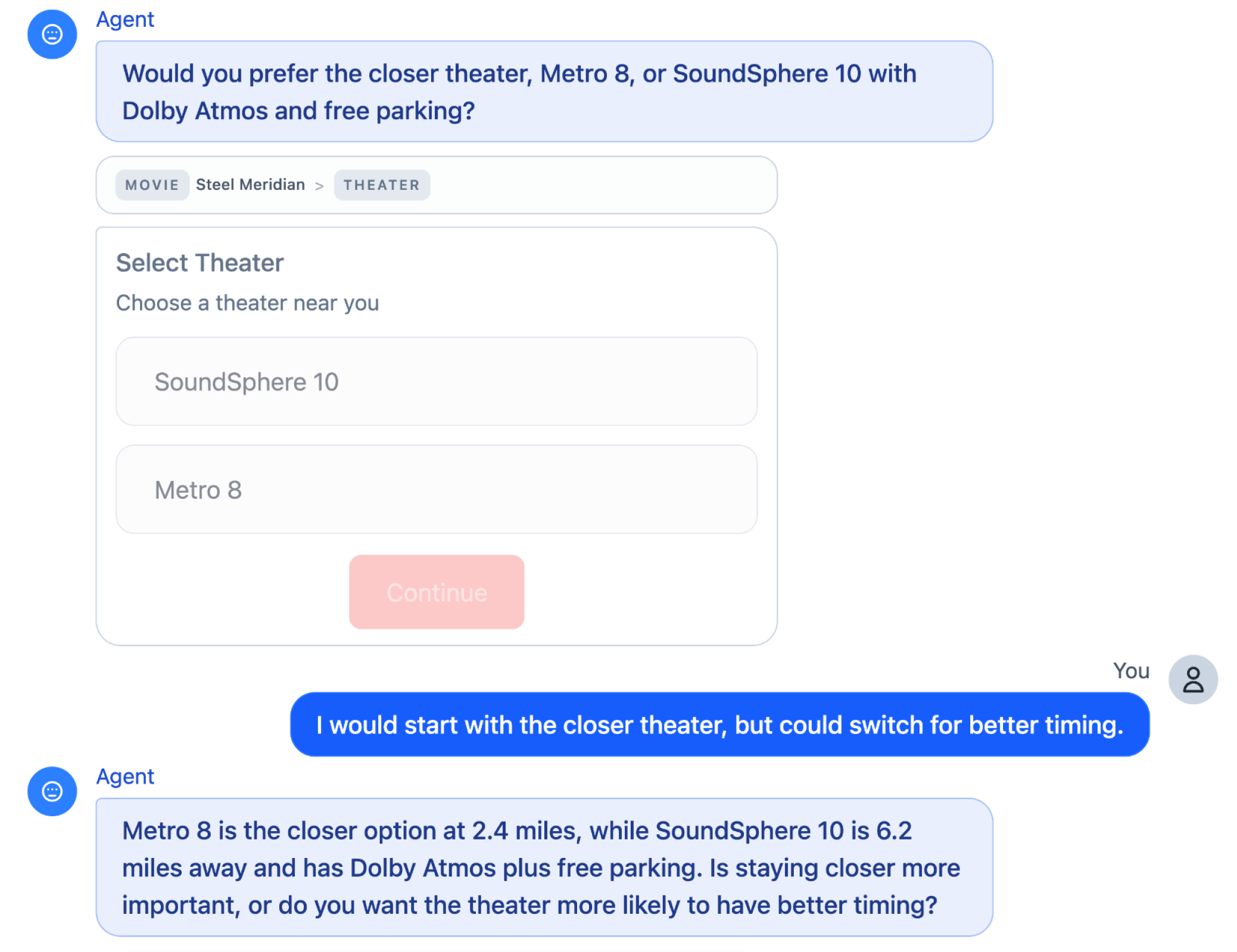}
  \caption{Baseline condition interface at the theater selection stage, designed after a typical CAG. A single chat-stream layout where GUI components and agent messages are mixed in a linear chat history. The baseline agent also has access to all metadata. 
  % timeline. \textbf{(A)}~A prior-stage GUI component (locked after selection). \textbf{(B)}~User's natural-language message expressing a preference. \textbf{(C)}~Agent's text-based response comparing options using backend metadata. \textbf{(D)}~The current stage's GUI component with selectable options and navigation buttons. \textbf{(E)}~Text input area.
  }
  \Description{A single-column chat interface for the Baseline condition at the theater selection stage. At the top, the agent asks whether the user prefers Metro 8 or SoundSphere 10. Below, a locked GUI component from a prior stage shows two theater buttons (SoundSphere 10 and Metro 8) with a Continue button. The user's message reads ``I would start with the closer theater, but could switch for better timing.'' The agent responds with a detailed text comparison: Metro 8 is 2.4 miles away and SoundSphere 10 is 6.2 miles with Dolby Atmos and free parking. Below, the current-stage GUI component shows the same two theater options with Back and Continue buttons. A text input bar appears at the bottom.}
  \label{fig:baseline-layout}
  \vspace{-10pt}
\end{figure}

In the \emph{Baseline condition} (\autoref{fig:baseline-layout}), participants used a single chat-stream layout common to most existing CAGs. The GUI component for the current stage was persistently displayed below the most recent message, and GUI components from previous stages remained visible in the chat history but were locked. Because no GUI adaptation tools were available, the screen did not change dynamically. In the \emph{\sys{} condition} (\autoref{fig:maestro-layout}), the screen was divided into a persistent GUI panel and a conversation panel, as described in \autoref{sec:target-domain}.

The baseline agent still had many useful features. Both conditions provide the agent with access to the same backend metadata; the difference lies solely in how the agent supports the user's decision process. 
Both conditions perform proactive preference elicitation---the agent actively asks about preferences upon entering a stage. In the Baseline, elicited preferences are used only within the conversation history; in \sys{}, they are stored as structured preference records.
\rev{The baseline agent thus represents a modern, GUI-integrated, task-oriented chatbot: its tools are limited to standard GUI interactions, namely selecting an option and moving to the next or previous stage, and in place of a structured memory module such as Preference Memory, it uses the conversation history as its only memory. That history covers the full session, whereas the \sys{} condition truncates it to recent turns (\autoref{sec:agent}); the system prompts of both conditions are reproduced in \autoref{sec:appendix-prompts}. Agents that persist structured records across sessions (e.g., ChatGPT's Memory) target long-term personalization and lie outside this comparison.}
%Both conditions perform proactive preference elicitation---the agent actively asks about preferences upon entering a stage. In the Baseline, these are used only within the conversation history; in \sys{}, they are stored as structured preference records.
%This elicitation is performed identically in both conditions: in the Baseline condition, it is used only within the conversation history, whereas in the \sys{} condition, the responses are stored as structured preference records.

% The UI separation in \sys{} is an architectural prerequisite for GUI Adaptation: in-place operators (highlight, sort, filter, augment) require a persistent GUI canvas. If the GUI is mixed into the chat stream, adaptations still take effect but are easily overlooked as modified elements scroll out of view, making it significantly harder for users to notice and act on the changes. UI separation is therefore part of the \sys{} design, not an independent feature. In the \sys{} condition, the GUI panel was interaction-locked during GUI adaptation to prevent accidental selections while the interface was being modified; the lock was also engaged while the agent was loading its next action.

\subsubsection{Mode: Text vs. Voice}

%We also explored two different modes of natural interaction with the CAG: Text vs. Voice to see how the effectiveness varies depending on the modality. 
The Mode factor varies the natural-language input/output channel: \textbf{Text} mode uses text typing plus GUI clicks with text-based agent responses, while \textbf{Voice} mode uses speech input plus GUI clicks with spoken agent responses and a text transcript. In Voice mode, the agent's spoken output was queued and played to completion before the input turn returned to the participant; once the participant activated the microphone, automatic turn-taking detected the end of the participant's utterance and triggered transcription. In both modes, participants can make simple selections via GUI clicks; the difference lies only in the natural-language channel. We used a commercial speech-to-text and text-to-speech engine to implement the Voice mode.

%% ------------------------------------------------------------
\subsection{Domain and Task Design}
\label{sec:domain}
%% ------------------------------------------------------------

\subsubsection{Domain: Movie Ticketing}

We chose movie ticketing as the study domain based on a content analysis of four major platforms (Fandango, AMC, Regal, Cinemark). All four share a common structure: Movie $\to$ Date/Theater/Time (compressed onto one screen) $\to$ Seat $\to$ Payment. We decomposed this into the six-stage linear workflow described in \autoref{sec:target-domain}. Because each stage's available options depend on selections made in earlier stages, this sequential structure forms the basis for cross-stage conflicts that \sys{}'s Workflow Navigation is designed to address.

\subsubsection{Task Design}

Eight tasks are organized into four task sets (T1--T4) to avoid learning effects from repeating the same task, with each set containing one warm-up and one main task. Warm-up tasks require 8 stage visits and 1 backtrack on the success path; main tasks require 18 visits and 2--3 backtracks. Warm-up tasks familiarise participants with the current condition and are excluded from the main analysis.

All tasks share the same interaction skeleton (stage order and UI types) but use different scenario backgrounds and scenario-specific databases.
Each task is presented as a scenario that includes a title, a brief background story, and a set of per-stage preference descriptions. Preferences are linguistically divided into \emph{hard} and \emph{soft}: hard preferences use obligatory language (e.g., ``I need a G or PG-rated movie''), while soft preferences use hedged language (e.g., ``I prefer the closest theater'' or ``preferably the shorter one''). 
Each scenario and paired data set has a single path that satisfies all the specified hard preferences. Example task prompts are shared in \autoref{sec:main_tasks}.
We typically used soft preferences to guide users toward a dead end and allowed them to use the system to return to previous steps to find a viable solution, indicating that no path satisfied both hard and soft preferences.
% In practice, users may dynamically relax preferences in response to real-world constraints; however, allowing such relaxation would make it difficult to define a unique correct workflow and to control experimental variables across conditions. Preferences were therefore held fixed throughout each task. 
% \sang{mention Appendix inclusion of Task screenshot and text}
%  Each task's database is designed so that exactly one end-to-end workflow satisfies all stated preferences while also meeting constraints revealed only through exploration. Each participant performs each pair exactly once, preventing content-learning effects that counterbalancing alone cannot offset.

% The scenarios adhere to the following design principles: (1)~each task has a unique feasible workflow; (2)~each uses a separate database; (3)~cross-stage conflicts are included so that Workflow Navigation's value is apparent; (4)~within-stage comparison is required so that GUI Adaptation's value is apparent; (5)~the stages at which conflicts and alternatives first emerge are distributed differently across pairs; and (6)~average complexity is balanced across the four pairs.

%% ------------------------------------------------------------
\subsection{Procedure}
\label{sec:procedure}
%% ------------------------------------------------------------

Each participant completed one task set (T1--T4) under each condition. 
Task--condition pairings and presentation order were fully counterbalanced using a Latin-square design (\autoref{tab:latin-square}), requiring $N$ to be a multiple of 16 ($4 \times 4$) to control for learning, order, and task--condition pairing effects.

%Participants were assigned to one of four groups (Groups 1–4), which determined the pairing between experimental conditions (BT, BV, MT, MB) and task sets (T1–T4). This ordering was counterbalanced using a Latin-square design.
%The task–condition grouping was rotated using a Latin-square design across participants. For example, the first four participants followed the group assignment shown in~\autoref{tab:latin-square}, and the next four participants used a different pairing (e.g., Groups 4, 5, 6, and 7). If the same pairings were used repeatedly, the BV condition would always be evaluated with T1, potentially introducing bias. Therefore, full counterbalancing requires the number of participants to be a multiple of 16 (4$\times$4).
%This design controls for learning and order effects, as well as potential bias arising from task–condition pairings.
% ; round and task–condition assignments are included as covariates in the analysis.

\begin{table}[t]
\caption{Latin-square Assignment of Task-Condition Group to Order for the first four participants.}
\Description{Table 1 shows the Latin-square assignment of task-condition groups to rounds for the first four participants. The table has five columns: Group, Round 1, Round 2, Round 3, and Round 4, and four rows for Groups 1 through 4. Each cell pairs a task set (T1 to T4) with a condition code, where B means Baseline, M means MAESTRO, T means Text, and V means Voice. Group 1 receives T1 with BT, T2 with MV, T3 with BV, and T4 with MT. Group 2 receives T2 with MT, T3 with BV, T4 with MV, and T1 with BT. Group 3 receives T3 with BV, T4 with MT, T1 with BT, and T2 with MV. Group 4 receives T4 with MV, T1 with BT, T2 with MT, and T3 with BV. Each task set and each condition appears exactly once per group, counterbalancing order across participants.}
\label{tab:latin-square}
\small
\begin{tabular}{lllll}
\toprule
\textbf{Group} & \textbf{Round 1} & \textbf{Round 2} & \textbf{Round 3} & \textbf{Round 4} \\
\midrule
Group 1 & T1\,·\,BT  & T2\,·\,MV & T3\,·\,BV & T4\,·\,MT \\
Group 2 & T2\,·\,MT   & T3\,·\,BV  & T4\,·\,MV & T1\,·\,BT \\
Group 3 & T3\,·\,BV  & T4\,·\,MT   & T1\,·\,BT & T2\,·\,MV\\
Group 4 & T4\,·\,MV  & T1\,·\,BT  & T2\,·\,MT  & T3\,·\,BV \\
\bottomrule
\end{tabular}
\vspace{-10pt}
\end{table}

Participants joined remotely from personal laptops. The protocol proceeded as follows: (1)~consent and pre-study survey; (2)~mandatory screen recording setup (webcam optional); (3)~tutorial via embedded video; (4--7)~four rounds, each consisting of a warm-up task followed by a main task and a per-cell survey; (8)~post-study survey with ranking and per-feature evaluation; and (9)~a semi-structured interview (${\sim}10$~minutes).

In each round, the warm-up task familiarised the participant with the current condition. During the warm-up, the researcher observed the session with their microphone and camera on and was available to answer questions about the system's use. The warm-up was occasionally skipped when the researcher judged the participant to be already sufficiently familiar with the condition. For the main task, the researcher turned off their microphone and camera and provided no assistance, allowing participants to work entirely on their own. Per-cell surveys were administered after each main task (4 times in total). The estimated session duration was approximately 90 minutes, and participants were compensated \$30.

Each task provides a scenario background and preference descriptions. Constraints are not stated in the brief but discovered through exploration. The brief remains on screen throughout.

%% ------------------------------------------------------------
\subsection{Dependent Variables}
\label{sec:measures}
%% ------------------------------------------------------------

\subsubsection{RQ1: Performance Measures}
To assess the performance of \sys{}, we developed objective measures as proxies for decision quality and efficiency. For decision-making quality, we compared their final submitted answer to the single solution that satisfies all the hard preferences in two ways: \emph{Task Success Rate} (DV1)---whether the final booking matches the unique correct workflow, 0 if failed, 1 if successful---and \emph{Violation Count} (DV2)---the number of hard preferences unmet; a lower value is better. We also measured \emph{Unpreferred Selection Count} (DV3) --- the total number of selected options that violate hard preferences across all stages within a task, which assesses the effectiveness of GUI adaptation during the process. For example, Unpreferred Selection Count can exceed Violation Count if a user repeats the same mistakes, as it captures errors made during the process, even if they are corrected later. \rev{Task Success Rate and Violation Count both assess the final outcome, with Violation Count grading the severity of failure that the binary success measure cannot capture, whereas Unpreferred Selection Count assesses the decision process leading to that outcome.} For Efficiency, we measured \emph{Task Completion Time} (DV4).

Another key factor in the success of CAGs is how users express their preferences. If users do not share their preferences and rely solely on GUI interactions, \sys{} cannot be effective. As we implemented proactive and encouraging language to prompt users to articulate their preferences, we measured how much users interacted in natural language and expressed preferences that cannot be inferred from GUI interactions alone.
We classified all user utterances, i.e., each message, into three functional categories using LLM-assisted coding validated against manual annotations: \emph{Preference Statement Utterances} (DV5) --- expressing a want, need, or priority (e.g., ``I prefer the closest theater''); \emph{Information-Seeking Utterances} (DV6) --- requesting factual information not visible on screen (e.g., ``How long is the movie?''); \emph{Action Request Utterances} (DV7) --- directing navigation or system actions (e.g., ``Go back'', ``March 11th''). We also report \emph{Total Utterances} (DV8).

% e.g., selecting non-adjacent seats when the task describes a preference of ``three seats together''), and unpreferred selection count (B3---the count of per-stage selections that fall outside the set of options consistent with the user's stated preferences and task constraints).

% \textbf{Efficiency.} Choice-ready-to-commit time (B4---the interval from when selectable options first appear on screen to when the participant commits a selection, excluding system preparation time), total step count including revisits with excess steps beyond 18 reported (B5), and backtrack count (B6, exploratory).

% \textbf{Utterance coding.} All user utterances were classified into four functional categories using LLM-assisted coding validated against manual annotations: \emph{preference statement}---expressing a want, need, or priority (e.g., ``I prefer the closest theater''); \emph{information-seeking}---requesting factual information not visible on screen (e.g., ``How long is the movie?''); \emph{action request}---directing navigation or system actions (e.g., ``Go back''); and \emph{other}---acknowledgements, confirmations, or off-topic remarks. The resulting counts and proportions per category (B7--B10) are reported alongside total utterance count (B11).

\subsubsection{RQ2: Perceived and Experiential Values}

% The subjective battery combines validated scales with custom items. 
We used a standard questionnaire to evaluate the perceived value of \sys{} compared to the baseline.
We measured Ease of Use using five items drawn from the Difficulty of Use and Mental Burden subscales of the \emph{User Burden Scale} (UBS)~\cite{suh2016developing}, averaging the five-point scale responses across these items.
To measure the Perceived Usefulness of \sys{}, we used four items from the \emph{Perceived Usefulness} (PU) subscale of the Technology Acceptance Model ~\cite{marangunic2015technology}, using a 7-point Likert scale (1 = Strongly disagree, 7 = Strongly agree).

% User Burden Scale (UBS) items use the original 5-point scale (0--4). Analysis defaults to individual-item reporting, with groupings reported as auxiliary composites when appropriate.

% \textbf{Custom---Comparison and Conflict} (3~items). These items target constructs unique to \sys{}: (C1)~ability to compare options within a stage, (C2)~understanding of how earlier preferences affect later stages, and (C3)~perception that the system proactively surfaces conflicts.

% : mental effort (U1), time taken (U2), learning difficulty (U3), information to remember (U4), and information overload (U5).

% \textbf{TAM PU---Perceived Usefulness} (4~items)..  assess perceived speed (PU1), effectiveness (PU2), ease (PU3), and usefulness (PU4).

% The per-cell survey burden is 12~items $\times$ 4~cells = 48~items total.

\emph{Post-study survey} \rev{was administered once after all four tasks. Participants ranked the four conditions (BT, BV, MT, MV) in order of preference and provided a rationale in open-ended responses.}
% (P2), and rate on 7-point Likert scales the helpfulness of GUI Adaptation (P3), Workflow Navigation (P4), Preference Memory (P5), and the separated UI layout (P6).

\emph{Exit interview} \rev{was administered at the end of the study and lasted approximately 10 minutes. The questions covered the most helpful features, confusing moments, relative preference between the two interface conditions, and suggestions for improvement.}
% (I2), cross-stage conflict awareness (I3), reactions to screen changes (I4), text-vs-voice naturalness (I5), perception of preference memory (I6), overall preference for the agent-assisted system vs.\ independent booking (I7), relative preference between the two interface conditions (I8), and suggestions for improvement (I9).

%% ------------------------------------------------------------
\subsection{Sample Size Justification \& Recruitment}
\label{sec:sample-size}
%% ------------------------------------------------------------

We needed $N = 32$ participants. The power analysis indicated that we need N=24 for a $2 \times 2$ within-subjects study with a medium effect size ($f = 0.25$), $\alpha = .05$, and power $= .80$. However, because the task-pair combinations and order groups were counterbalanced, we needed the number of participants to be a multiple of 16. Considering attrition, we recruited $36$ participants in total. \rev{Two participants' data were excluded because technical issues (e.g., OpenAI API errors, system errors) disrupted their sessions, and one participant who failed all four main tasks was excluded from the analysis as an outlier, yielding $N=33$.}
% The G*Power estimate is based on per-participant cell averages; analyses using finer-grained task- or stage-level data were therefore treated as supplementary or exploratory. Interaction effects were reported exploratorily, given their lower statistical power.

We recruited participants through various mailing lists involving students at the authors' university. All participants (13 female, 20 male) were undergraduate or graduate students. Participants' ages were distributed as follows: 18–24 (n = 17, \rev{51.52\%}), 25–34 (n = 13, \rev{39.39\%}), and 35–44 (n = 3, \rev{9.09\%}). The study lasted approximately 1.5 hours. All participants provided informed consent and were compensated \$30 for their participation. The experimental protocol was reviewed and approved by the Institutional Review Board at our university. 
% \sang{This paragraph needs to be completed. }

%% ------------------------------------------------------------
\subsection{Analysis}
\label{sec:analysis-plan}
%% ------------------------------------------------------------

% Only the main tasks were included in the main analysis. All $N=32$ participants contributed data to every cell, yielding $128$ main-task trials with no exclusions.

The primary analysis uses mixed-effects models with participant as a random intercept and Condition, Mode, and their interaction as fixed effects. The model family is chosen based on outcome type: binomial GLMM for binary outcomes (DV1), Poisson GLMM for count outcomes (DV2, DV3, DV5--DV8), and Gaussian LMM applied to log-transformed time values (DV4).
For the composite scores of survey responses (UBS and PU), we analyzed them using a linear mixed-effects model.
% For all quantaitive measure, we also 
Open-ended responses and semi-structured interview transcripts were analyzed using open coding; two researchers independently coded responses and consolidated codes through discussion until they agreed. 

% ---for counts with overdispersion (e.g., requirement violations, stage visits), continuous measures (e.g., deliberation time), and Likert-scale survey items. The Condition~$\times$~Mode interaction is explored through pairwise cell comparisons (e.g., MT vs.\ BT), each fitted as a separate sub-model on the relevant two-cell subset.

% Post-study rankings (P1) were analysed with a Friedman test followed by pairwise Wilcoxon signed-rank tests. Post-study direct-evaluation items (e.g., P3, P4) were analysed with one-sample Wilcoxon signed-rank tests against the scale midpoint.

% To evaluate the accuracy of \sys{}'s workflow navigation, each backtrack suggestion was compared against a \textit{gold oracle}---a reference trace pre-computed for every reachable state in each task scenario that records the correct backtrack target and the number of viable alternatives at each prior stage.

%% ============================================================
\section{Results}
\label{sec:results}
%% ============================================================

All analyses use main-task data only \rev{($N=33$, $132$ trials; one participant was excluded as an outlier, see \autoref{sec:sample-size})}. The statistical approach follows Section~\ref{sec:analysis-plan}; all $p$-values are two-sided and unadjusted unless stated otherwise.

\begin{figure*}[t]
  \centering
  \includegraphics[width=\textwidth, trim=0 12.7bp 0 7.2bp, clip]{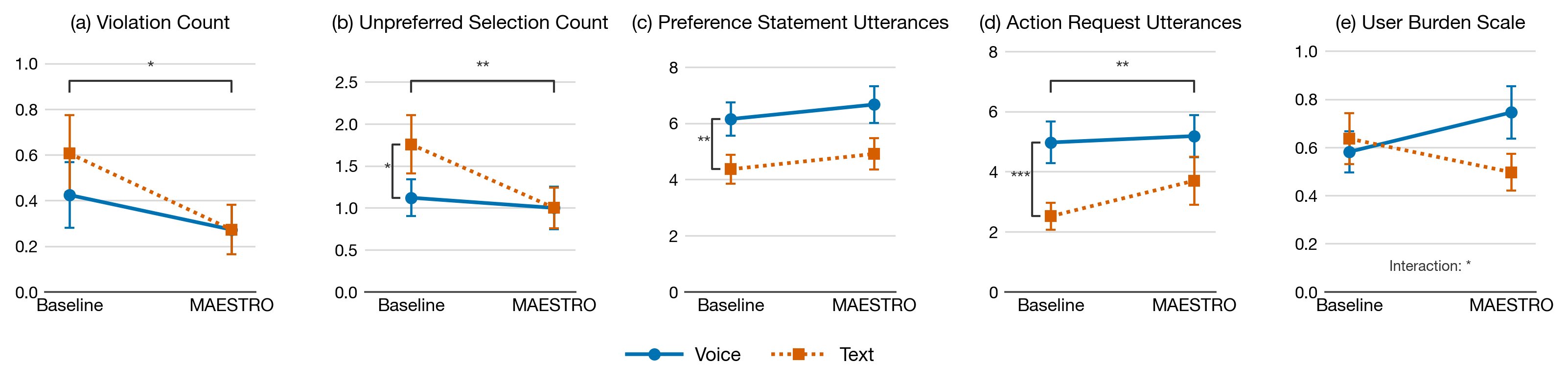}
  \vspace{-10pt}
  \caption{Interaction plots for five dependent variables used in the evaluation: Violation Count (a), Unpreferred Selection Count (b), Preference Statement Utterances (c), Action Request Utterances (d), and User Burden Scale (e). Each panel compares Baseline and \sys{} with separate lines for Voice and Text. Error bars denote $\pm 1$~standard error (SE). Statistical significance is annotated directly on each panel: top horizontal brackets indicate a significant Condition effect (Baseline vs.\ \sys{}), left-side brackets indicate a significant Mode effect (Voice vs.\ Text), and ``Interaction: *'' indicates a significant Condition $\times$ Mode interaction. Asterisks follow the standard convention (* $p<.05$, ** $p<.01$, *** $p<.001$).}
  \Description{A wide five-panel figure summarizing key dependent variables from the study. Panel (a), Violation Count, shows lower counts under MAESTRO than Baseline in both voice and text. Panel (b), Unpreferred Selection Count, shows fewer unpreferred selections under MAESTRO and a higher count in text than voice at Baseline. Panel (c), Preference Statement Utterances, shows more preference-statement utterances in voice than text, with both increasing slightly under MAESTRO. Panel (d), Action Request Utterances, shows more action-request utterances in voice than text and an increase in text under MAESTRO. Panel (e), User Burden Scale, shows an interaction pattern in which voice increases under MAESTRO while text decreases. All panels use blue solid lines with circle markers for Voice and orange dotted lines with square markers for Text, with error bars and significance annotations.}
  \label{fig:fiveplots}
  \vspace{-6pt}
\end{figure*}

% \begin{figure}[t]
%     \centering
%     \begin{subfigure}{0.19\textwidth}
%         \includegraphics[width=\linewidth]{figures/graphs/violation.png}
%         \caption{(a)}
%     \end{subfigure}
%     \begin{subfigure}{0.19\textwidth}
%         \includegraphics[width=\linewidth]{figures/graphs/unpreferred_selection_count.png}
%         \caption{(b)}
%     \end{subfigure}
%     \begin{subfigure}{0.19\textwidth}
%         \includegraphics[width=\linewidth]{figures/graphs/preference_statement_count_llm.png}
%         \caption{(c)}
%     \end{subfigure}
%     \begin{subfigure}{0.19\textwidth}
%         \includegraphics[width=\linewidth]{figures/graphs/action_request_count_llm.png}
%         \caption{(d)}
%     \end{subfigure}
%     \begin{subfigure}{0.19\textwidth}
%         \includegraphics[width=\linewidth]{figures/graphs/survey_ubs_mean.png}
%         \caption{(e)}
%     \end{subfigure}
%     \caption{Your overall caption describing all five graphs.}
%     \label{fig:five_plots}
% \end{figure*}

%% ------------------------------------------------------------
\subsection{RQ1: How \sys{} supported decision making}
\label{sec:results-rq1}
%% ------------------------------------------------------------
\subsubsection{\texorpdfstring{\rev{Performance Metrics: \sys{} reduces preference violations.}}{Performance Metrics: MAESTRO reduces preference violations.}}
\rev{Overall, \sys{} improved two specific aspects of decision quality: the final outcome, measured by Violation Count, and the decision process leading to it, measured by Unpreferred Selection Count.} While the difference in Task Success Rate (DV1) was not statistically significant, we observed reductions in \rev{these two preference-violation measures}, which can serve as reverse proxies for decision quality. \rev{One possible explanation for the absence of a significant difference in success rate is that the binary measure is insensitive to partial improvements: a booking that violates fewer, but not zero, hard preferences still counts as a failure.}
\rev{Descriptive statistics for all dependent variables, including those without significant effects, are provided in \autoref{tab:appendix-dvs}.}

Violation Count was lower (DV2) in the \sys{} condition, and the difference was statistically significant. There was a significant main effect of condition, with \sys{} reducing violation counts compared to the baseline ($\beta$ = -0.80, SE = 0.40, $z$ = -1.99, $p$ = .047). Neither the main effect of mode nor the interaction effect was significant.
This result suggests that the tickets selected using \sys{} had fewer unmet hard preferences, demonstrating its contribution to improving decision quality. The results for Violation Count are shown in \autoref{fig:fiveplots}-(a). 
% \sang{make sure to include sub labels of figures}

Unpreferred Selection Count (DV3) was also reduced in the \sys{} condition, and the difference was statistically significant. There was a significant main effect of condition, with \sys{} reducing unpreferred selection counts compared to the baseline ($\beta$ = -0.56, SE = 0.21, $z$ = -2.64, $p$ = .008). We also found a significant main effect of mode, with Voice reducing unpreferred selection counts compared to Text ($\beta$ = -0.45, SE = 0.21, $z$ = -2.18, $p$ = .029). The interaction effect was not significant.
This result suggests that \sys{} helps users avoid selecting options that violate hard preferences during decision-making, thereby improving decision quality, as reflected in DV2. The results for Unpreferred Selection Count are shown in \autoref{fig:fiveplots}-(b). 

While overall Task Completion Time \rev{(DV4)} was longer for \sys{}, the difference was not statistically significant. In practice, we observed that time spent adapting the GUI contributed to delays in the \sys{} condition. \rev{Although this trend suggests a modest time cost, we found no evidence that supporting the decision process significantly slowed task completion.}

\subsubsection{Utterance Patterns: \sys{} encourages people to interact with the GUI in natural language. }

We found meaningful differences in how they interact with \sys{} compared to the baseline system. 
Preference Statement Count (DV5) was higher in the Voice mode, and the difference was statistically significant. There was a significant main effect of mode, with Voice increasing preference statement counts compared to Text ($\beta$ = 0.34, SE = 0.11, $z$ = 3.18, $p$ = .002). The main effect of Condition was not significant,
%($\beta$ = 0.12, SE = 0.11, $z$ = 1.04, $p$ = .300)
nor was the interaction effect. The results for Preference Statement Count are shown in \autoref{fig:fiveplots}-(c). 
% ($\beta$ = -0.04, SE = 0.15, $z$ = -0.25, $p$ = .802).
This result suggests that users expressed their preferences more frequently when interacting via voice, while the \sys{} condition had no significant effect. 
% Given that more expressed preferences provide richer input to the system, this tendency may lead to greater opportunities for \sys{} to support decision-making, although this effect was not directly observed this time.

Action Request Count (DV7) was higher in both the \sys{} condition and the Voice mode. There was a significant main effect of condition, with \sys{} increasing action request counts compared to the baseline ($\beta$ = 0.39, SE = 0.14, $z$ = 2.74, $p$ = .006). We also found a significant main effect of mode, with Voice increasing action request counts compared to Text ($\beta$ = 0.68, SE = 0.13, $z$ = 5.11, $p$ < .001). The interaction effect was marginal but not statistically significant ($\beta$ = -0.34, SE = 0.18, $z$ = -1.94, $p$ = .053). The results for Action Request Count are shown in \autoref{fig:fiveplots}-(d). 
Information-seeking utterances (DV6) did not show statistically significant effects for either factor. \rev{Total Utterance Count (DV8) was higher in both the \sys{} condition ($\beta$ = 0.17, SE = 0.08, $z$ = 2.28, $p$ = .023) and the Voice mode ($\beta$ = 0.41, SE = 0.07, $z$ = 5.69, $p$ < .001), with no significant interaction.}
Overall, these results suggest that \sys{} and voice interaction \rev{encourages more active and expressive interaction; the additional utterances under \sys{} were driven primarily by action requests}. \rev{Example conversations from both modes are provided in \autoref{sec:appendix-conversations}.}

\subsection{RQ2: The Perceived Value of \sys{}}
\label{sec:results-rq2}
%% ------------------------------------------------------------

\subsubsection{More people preferred \sys{} over the baseline agent}

We measured subjective responses with Perceived Usefulness (PU) and the User Burden Scale (UBS). 
For Perceived Usefulness (PU), we found no evidence that \sys{} was perceived as more useful than the baseline.
% One possible explanation is that the baseline agent was already sufficiently capable of addressing users’ needs, leading to uniformly high evaluations across conditions. This ceiling effect may have limited our ability to detect differences in perceived usefulness between the two systems.
However, the exit survey clearly showed that participants preferred \sys{} over the Baseline agent. 

When asked to rank all four conditions after completing the study, participants showed a clear preference for \sys{}. The Friedman test was significant ($\chi^2(3)=22.45$, $p<.001$). Pairwise Wilcoxon signed-rank tests showed that \sys{} was preferred over Baseline in both Text mode (MT vs.~BT: $\Delta=-1.03$, $p<.001$) and Voice mode (MV vs.~BV: $\Delta=-0.48$, $p=.028$). Mean ranks placed \sys{} Text as the most preferred condition (MT: $1.64$), followed by \sys{} Voice (MV: $2.61$), Baseline Text (BT: $2.67$), and Baseline Voice (BV: $3.09$). Consistent with the ranking results, first-choice selections also favored \sys{}. A majority of participants selected \sys{} Text (MT) as their top choice (19 participants), followed by \sys{} Voice (MV; 7), Baseline Text (BT; 5), and Baseline Voice (BV; 2).

In open-ended responses, participants reported several reasons for preferring \sys{} over the Baseline agent. Many highlighted the side-by-side interface as more intuitive, less cluttered, and more interactive. Participants also valued \sys{}'s ability to retain preferences when backtracking, which helped them resume decision-making without restarting the process. Additionally, direct manipulation of the GUI (e.g., filtering, sorting, and augmenting information) was perceived to reduce cognitive load compared to reading text-based responses. Several participants noted that \sys{} provided a greater sense of control by supporting both GUI interaction and conversational input.

However, a subset of participants preferred the Baseline due to its familiar chat-based interaction style and perceived speed. While \sys{}'s transparent actions (e.g., showing filtering and sorting operations) were appreciated for improving system understanding and trust, they were also associated with increased latency, occasional feelings of restriction, and sensory overload.

\subsubsection{\sys{}'s verbosity can amplify a user's burden}
We calculated Mental Burden and Difficulty of Use from the User Burden Scale (UBS) as reverse proxies for Ease of Use. UBS did not show significant main effects of Condition or Mode. However, we found a significant interaction between condition and mode ($\beta$ = 0.30, SE = 0.14, $t$ = 2.20, $p$ = .030). As shown in \autoref{fig:fiveplots}-(e), users’ burden was higher when using \sys{} in Voice mode, whereas the opposite pattern was observed in Text mode.

We identified a strong and relevant pattern in the qualitative results that helps explain this finding. The turn-taking nature of the implemented Voice mode—where users could not interrupt while the agent was speaking—led to frustration. Many participants reported annoyance due to latency and speech recognition inaccuracies. The following quotes illustrate these experiences.

\begin{itemize}[leftmargin=1.5em]
    \item (P4) \textit{``I felt that the narrator was reading too much of the script, and at certain moments, I felt that, uh, I need to wait for the agent to complete its script before I could give my feedback.''}
    \item (P5) \textit{``As someone who is multilingual, it makes me mad when what I am saying is not transcribed accurately.''}
    \item (P25) \textit{In the voice interface, there was a bit of lag, and I kept talking over the AI assistant.}
\end{itemize}

% (P5) 
% As someone who is multilingual, it makes me mad when what I am saying is not transcribed accurately, so I don't like using the microphone to do tasks that I can accomplish using text/typing. (P5)

% I mentioned this in my post study survey as well. I do not like it when um, I'm saying something in the microphone and they transcribe it incorrectly, even if, even if um, the agent still understands what I tried to say. Like even if the job is done, if it is transcribing me incorrectly, it just pisses me off.
% another thing that I don't like is when the chatbot, um, keeps talking. Right? And so I wanted to be as precise as possible. Um, in speech. Right. Because I, I noticed that I had moved past that stage right to another stage where I was selecting the seeds or where I was selecting, um, the time, for instance, and I was still speaking about the movie theater choice.

% in the voice interface, there was a bit of lag, and I kept talking over the AI assistant. (P25) 

% ther than that, uh, yeah, I think, uh, in the voice, uh, voice one, uh, the…I felt that the narrator was reading too much of the script And at certain moments, I felt that, uh, I need to wait for the agent to complete its script before I could give my feedback. (P4) 

% 
\noindent Because speech recognition was identical across the conditions, the inherent verbosity of \sys{} may have amplified this frustration and contributed to the UBS interaction effect. \sys{} was designed to provide continuous feedback, and its GUI adaptation introduced additional messages whenever the interface was updated, potentially increasing perceived wait time.

To examine this, we analyzed the number of agent utterances. Agent Utterance Count was higher in the \sys{} condition. There was a significant main effect of Condition, with \sys{} producing more utterances than the baseline ($\beta$ = 0.51, SE = 0.04, $z$ = 12.70, $p$ < .001), yielding, on average, 20 more messages per task.
These results suggest that \sys{} generates more system feedback during interaction, which may contribute to increased perceived burden, particularly in Voice mode, where turn-taking delays further accumulate.

\subsubsection{Towards Agentic Exploration}

% One divisive feedback that we received from the participants was people's desire for the agent to have more autonomy vs for themselves to have agency in what they are doing by themselves. 

One common form of feedback we received from participants, as well as a recurring pattern observed during the tasks, was the expectation that \sys{} could perform forward search. For example, multiple users asked questions such as, “I would like a showtime that has three adjacent seats,” when prompted to select a showtime. However, \sys{} does not have the ability to look ahead to future stages; it only logs interaction traces and preferences based on past actions. The following comments illustrate this expectation.
\begin{itemize}[leftmargin=1.5em]
    \item (P2) \textit{``Letting us know if there's premium seating, or seating preferences while we're choosing the timing or the theater, I think, it would make the process a little faster.''}
    \item (P24) \textit{`` I kind of hoped that it would be able to see information like on the next step. it'll be nice if I could just write all my preferences down and like lists, like the available options from there.''}
    % \item (P5) \textit{``As someone who is multilingual, it makes me mad when what I am saying is not transcribed accurately.''}
    % \item (P25) \textit{In the voice interface, there was a bit of lag, and I kept talking over the AI assistant.}
\end{itemize}

\noindent  While this is not technically infeasible, it would require substantial computational resources to exhaustively search the entire solution space without human involvement, potentially relying on brute-force methods. These results suggest that users may form expectations that the agent can effectively explore the full search space on their behalf.

\section{Discussion}
\label{sec:discussion}
%% ============================================================
We found that \sys{} \rev{reduces preference violations in both the final outcome and the decision process, without a statistically significant increase in task completion time}. These benefits come with trade-offs, particularly in Voice mode, where increased feedback and latency originating from intelligent adaptation increase perceived burden, which future design should consider.

\subsection{Understanding Users More Deeply through Decision Support}

\sys{} improved decision quality by adapting the GUI to reflect users' preferences. Beyond these immediate benefits, an important opportunity lies in leveraging accumulated preferences over time to develop a deeper understanding of users. Prior work has explored representing users' long-term preferences as structured propositions to personalize LLM outputs and support agentic behavior~\cite{shaikh2025creatinga}, as well as maintaining persistent user memory~\cite{bae2022keep}. 

In particular, the ways users compromise and adjust their priorities through this interaction can reveal more nuanced aspects of their lifestyles or tastes, especially when preferences are revised in response to conflicts. For example, some users may prioritize a specific movie and remain committed to it despite constraints, whereas others may treat such preferences more flexibly, adjusting them when conflicts arise. These patterns suggest that decision-support systems like \sys{} can move beyond immediate task assistance toward modeling users' underlying priorities over time.

This type of nuanced understanding can be valuable for improving the perceived quality of web automation agents that must autonomously make decisions without user interruption. Current approaches often rely on human-in-the-loop mechanisms~\cite{huq2025cowpilot, peng2025morae, chen2023miwa} in specific contexts, such as accessibility. However, these moments of human intervention may be precisely where deeper reasoning about users’ multiple, potentially conflicting preferences is required. Enabling agents to better model and coordinate such preferences could enhance their ability to act more effectively on behalf of users.
\subsection{Rethinking Voice Interaction for Assisting User In Agentic Ways}

We observed two ambivalent aspects of voice mode in its support for decision-making. On one hand, using CAGs in voice mode increased certain types of utterances, particularly action requests expressed in natural language rather than through GUI interaction, as well as preference statements. This provides users with more opportunities to express their intentions and preferences. On the other hand, voice interaction increased users' perceived burden, especially because it delivers system feedback through the same primary modality used for GUI adaptation.

We believe that improving voice interaction in a more natural and efficient manner presents a promising direction for future design. For instance, not all feedback needs to be conveyed through voice; non-verbal feedback—particularly visual cues—can effectively communicate system actions. For example, the animation of the Sort operator clearly illustrates how the system reorganizes information to support decision-making.

In addition, modern voice-to-voice APIs support more natural interaction by allowing users to interrupt the agent while it is speaking. In such cases, the agent need not halt its operations; it can continue adapting the GUI in the background while processing user input, enabling the agent's multitasking capabilities.

\begin{revblock}
\subsection{Calibrating Agent Initiative}
\label{sec:agency}
%% ------------------------------------------------------------

Our findings also speak to how much initiative a task-oriented CAG should take. Participants welcomed initiative that stayed within the current stage, citing the adaptation operators as reasons for preferring \sys{}. Several participants expected more, asking the agent to look ahead to later stages and pre-screen options on their behalf (\autoref{sec:results-rq2}), a level of planning that \sys{} does not attempt. However, initiative carries a communication cost that the interaction modality must absorb, as the same proactive feedback that was acceptable in text amplified perceived burden in voice. These observations suggest calibrating initiative to the feedback channel and to the reversibility of the action: propose and confirm before committing on the user's behalf, keep feedback visual where possible in voice mode, and reserve autonomous execution for steps the user can easily undo.
\end{revblock}

%% ------------------------------------------------------------
\begin{revblock}
\subsection{Generalizing \sys{} Beyond Movie Ticketing}
\label{sec:generalizability}
%% ------------------------------------------------------------

Although we instantiated \sys{} in a movie-ticketing workflow, its mechanisms operate on a generic stage and attribute schema rather than on domain-specific logic. They can therefore extend to tasks that proceed through selection stages built from standard GUI widgets (e.g., drop-down menus, checkboxes, and date pickers), such as flight booking and step-by-step troubleshooting. Adapting \sys{} to a new domain involves coding the stages, describing each stage's attributes for the adaptation operators, and adjusting the prompt that assigns preferences to relevant stages.

Richer domains raise two design questions: how to prioritize the attributes that the augment operator surfaces when layout space is limited, and how to support flexible stage orders, as when a user picks a theater before a movie. The agent could treat each ordering as a distinct linear workflow selected at the outset, much as it selects tools, leaving the navigation mechanism unchanged. In the long term, because adaptation occurs in place, \sys{} could run as a browser extension that modifies the markup of existing web applications, extending preference-grounded adaptation to tasks such as online shopping without redesigning the underlying services.
\end{revblock}

%% ------------------------------------------------------------
\subsection{Limitations}
\label{sec:limitations}
%% ------------------------------------------------------------

Several limitations should be noted. First, our study uses a single domain---movie ticketing---and generalisability to more complex or heterogeneous domains remains to be established. Second, the comparison is between Baseline and \sys{} as a bundle (separated UI + GUI Adaptation + Workflow Navigation), so the independent causal contributions are not isolated. We provide indirect evidence through RQ-specific measure patterns, but rigorous factorial decomposition is left for future work. Third, the remote study setting limits control over the physical environment and introduces variability in participants' hardware and connectivity. Fourth, the fixed-preference design, while necessary for experimental control, does not capture how users might dynamically relax their preferences in response to real-world constraints. \rev{Fifth, we did not evaluate the accuracy of Preference Memory in isolation, as extraction performance depends not only on the system but also on whether and how users articulate their preferences; we instead refined extraction reliability through iterative testing during the pilot study, and a technical evaluation remains future work.}

% Several directions merit further investigation. Extending preference capture beyond language to incorporate behavioral signals---as suggested by the silent preference problem---is a natural next step. The current set of four adaptation operators may need to be extended for domains with richer GUI structures. The voice--latency tension calls for research on interruptible and progressive adaptation strategies. Finally, the Shared Preference Memory is currently limited to within-session operation; extending it to long-term preference learning or cross-domain preference transfer presents promising opportunities.

\bibliographystyle{ACM-Reference-Format}
\bibliography{references}

\appendix

%% ============================================================
%% Appendix (split out of main.tex for the camera-ready)
%% ============================================================
\section{Appendix}
\subsection{Sample JSON Object in Preference Memory}
\label{sec:JSON}

\begin{lstlisting}[style=json]
[
    {
        "description": "comedy movie",
        "strength": "soft",
        "relevantStages": ["movie"]
    },
    {
        "description": "Friend A must be home by 10 PM",
        "strength": "hard",
        "relevantStages": ["time", "seat"]
    },
    {
        "description": "prefer higher-rated movies",
        "strength": "soft",
        "relevantStages": ["movie"]
    }
]
\end{lstlisting}

\subsection{Example Scenarios for Main Tasks}
\label{sec:main_tasks}
\subsubsection{Parents Anniversary Gift}

\begin{description}[leftmargin=0.4cm,style=nextline]
\item[Background.] I am trying to set up a movie outing for my parents' anniversary weekend, and I want it to feel warm, comfortable, and a little special.
\item[Movie stage.] I need a PG-13 or below romance movie, preferably warm and familiar in tone.
\item[Theater stage.] I would start with the closer theater, but can switch for better timing or seating.
\item[Date stage.] I prefer Saturday, March 14, but Sunday, March 15 also works.
\item[Showtime stage.] I need it to start after 4:00 PM and end by 9:00 PM on Saturday, and on Sunday I need a true morning show.
\item[Seat stage.] I need two adjacent premium seats---standard would feel too ordinary.
\end{description}

\subsubsection{Sibling B-Movie Comedy Night}

\begin{description}[leftmargin=0.4cm,style=nextline]
\item[Background.] I am planning a movie night with my sibling, and I want it to feel weird and fun in exactly the right low-budget way, even if I have to try a couple of paths first.
\item[Movie stage.] I want a cult comedy, and I prefer the lower-rated one over the higher-rated one.
\item[Theater stage.] I need it at the single-screen theater.
\item[Date stage.] I can go on Friday, March 13 or Saturday, March 14, and I prefer Friday.
\item[Showtime stage.] I need it starting after 6:00 PM and ending by 10:00 PM, preferring the earlier showtime.
\item[Seat stage.] I need two adjacent seats, not in the back rows.
\end{description}

\begin{revblock}
\subsection{Example Conversations by Mode}
\label{sec:appendix-conversations}
To illustrate how interaction differed between modes, we reproduce excerpts from two \sys{} sessions on the same task (Sibling B-Movie Comedy Night, \autoref{sec:main_tasks}). In these excerpts, the Text-mode participant states preferences compactly and confirms quickly, whereas the Voice-mode participant speaks in longer, hedged turns, requests summaries, and asks the agent for a suggestion.

\medskip
\noindent\textbf{\sys{} in Text mode (P32; movie and theater stages).}
\begin{itemize}[label={}, leftmargin=1em, itemsep=1pt]
  \item \textbf{Agent:} What kind of movie are you in the mood for?
  \item \textbf{P32:} I want a cult comedy and I prefer the lower-rated one over the higher-rated one.
  \item \textbf{Agent:} Showing genre and rating so your movie preferences are easier to compare.
  \item \textbf{Agent:} Showing the cult comedies and favoring the lower-rated one among them.
  \item \textbf{Agent:} The lower-rated cult comedy here is Slime Beach Weekend. Want to go with that one?
  \item \textbf{P32:} yes
  \item \textbf{Agent:} Got it --- Slime Beach Weekend.
  \item \emph{(theater stage)}
  \item \textbf{Agent:} Which theater works best for you?
  \item \textbf{P32:} Which one has single-screen?
  \item \textbf{Agent:} Showing screen counts for these theaters.
  \item \textbf{Agent:} Showing theaters with 1 screen.
  \item \textbf{Agent:} Picture Shack is the single-screen theater here. Want to go with that?
  \item \textbf{P32:} yes
\end{itemize}

\medskip
\noindent\textbf{\sys{} in Voice mode (P01; movie stage).}
\begin{itemize}[label={}, leftmargin=1em, itemsep=1pt]
  \item \textbf{Agent:} What kind of movie are you in the mood for today?
  \item \textbf{P01:} I am interested for a movie night with my sibling. Um, I want it to feel weird and fun. What would you suggest?
  \item \textbf{Agent:} I'm showing which movies lean more weird and fun.
  \item \textbf{Agent:} I'm highlighting the movies that best match weird and fun.
  \item \textbf{Agent:} The best matches for that vibe are Slime Beach Weekend and Motel of Miracles. Which one sounds better to you?
  \item \textbf{P01:} Tell me a small summary between the two, like two to three sentences.
  \item \textbf{Agent:} Here are short summaries for the two highlighted movies.
  \item \textbf{Agent:} The short summaries are on screen now for those two picks. Which one fits your movie night better?
  \item \textbf{P01:} I would prefer like a rougher B-movie or cult comedy.
  \item \textbf{Agent:} I'm highlighting the cult-comedy picks that fit that rougher B-movie vibe.
  \item \textbf{Agent:} Both highlighted movies fit that rougher cult-comedy vibe. Which one do you want to go with: Slime Beach Weekend or Motel of Miracles?
  \item \textbf{P01:} Not a prestige one. Which one do you suggest?
  \item \textbf{Agent:} Slime Beach Weekend sounds like the better fit for that rougher B-movie, cult-comedy vibe. Want to go with that one?
  \item \textbf{P01:} Yes.
\end{itemize}
\end{revblock}

\clearpage
\onecolumn
% ============================================================
% Appendix DV table (rebuttal promise to 2AC). Values computed from
% analysis/results/final/raw_data.csv — final analysis cohort N=33
% (34 completed participants minus one outlier who failed all four
% tasks; P50/P51 included). Cell entries are mean (SD), SD with ddof=1.
% Cross-checked against the reported mixed-effects statistics (which
% reproduce exactly on this cohort).
% ============================================================
\begin{revblock}
\subsection{Descriptive Statistics for All Dependent Variables}
\label{sec:appendix-dvs}
\autoref{tab:appendix-dvs} reports descriptive statistics for all dependent variables across the four conditions, including measures that did not show statistically significant effects.

\begin{table*}[!ht]
  \caption{Descriptive statistics for all dependent variables by condition (mean (SD); Task Success Rate as the percentage of successful trials). BT/BV denote Baseline in Text/Voice mode; MT/MV denote \sys{} in Text/Voice mode. The Sig.\ column marks statistically significant effects from the mixed-effects models (\autoref{sec:analysis-plan}): C = Condition main effect (Baseline vs.\ \sys{}, i.e., BT+BV vs.\ MT+MV), M = Mode main effect (Text vs.\ Voice, i.e., BT+MT vs.\ BV+MV), C$\times$M = their interaction (* $p<.05$, ** $p<.01$, *** $p<.001$).}
  \Description{Table 2 reports descriptive statistics for all eleven dependent variables across the four conditions, given as mean with standard deviation in parentheses. The six columns are Measure, BT (Baseline Text), BV (Baseline Voice), MT (\sys{} Text), MV (\sys{} Voice), and Sig., which marks significant effects from the mixed-effects models, where C is a Condition main effect, M is a Mode main effect, and C by M is their interaction. Task Success Rate is 61\%, 67\%, 76\%, and 76\%, with no significant effect. Violation Count is 0.61 (0.97), 0.42 (0.83), 0.27 (0.63), and 0.27 (0.63), with a significant Condition effect. Unpreferred Selection Count is 1.76 (2.00), 1.12 (1.27), 1.00 (1.39), and 1.00 (1.46), with significant Condition and Mode effects. Task Completion Time in seconds is 349.2 (154.1), 406.0 (142.3), 403.1 (197.8), and 466.9 (182.0), with no significant effect. Preference Statement Utterances is 4.36 (2.92), 6.15 (3.37), 4.91 (3.21), and 6.67 (3.78), with a significant Mode effect. Information-Seeking Utterances is 2.09 (1.97), 2.36 (2.09), 1.82 (1.86), and 1.88 (1.65), with no significant effect. Action Request Utterances is 2.52 (2.60), 4.97 (3.96), 3.70 (4.58), and 5.18 (4.02), with significant Condition and Mode effects. Total Utterances is 9.64 (6.34), 14.52 (7.26), 11.45 (8.33), and 15.21 (8.56), with significant Condition and Mode effects. Agent Utterance Count is 29.6 (12.0), 36.5 (13.6), 49.3 (21.0), and 55.8 (19.9), with significant Condition and Mode effects. User Burden Scale on a 0 to 4 range is 0.64 (0.60), 0.58 (0.49), 0.50 (0.44), and 0.75 (0.63), with a significant Condition by Mode interaction. Perceived Usefulness on a 1 to 7 scale is 5.70 (1.11), 5.46 (1.26), 5.73 (1.21), and 5.44 (1.33), with no significant effect.}
  \label{tab:appendix-dvs}
  \small
  \begin{tabular}{lrrrrl}
  \toprule
  \textbf{Measure} & \textbf{BT} & \textbf{BV} & \textbf{MT} & \textbf{MV} & \textbf{Sig.} \\
  \midrule
  Task Success Rate (DV1) & 61\% & 67\% & 76\% & 76\% & --- \\
  Violation Count (DV2) & 0.61 (0.97) & 0.42 (0.83) & 0.27 (0.63) & 0.27 (0.63) & C* \\
  Unpreferred Selection Count (DV3) & 1.76 (2.00) & 1.12 (1.27) & 1.00 (1.39) & 1.00 (1.46) & C**, M* \\
  Task Completion Time (DV4, s) & 349.2 (154.1) & 406.0 (142.3) & 403.1 (197.8) & 466.9 (182.0) & --- \\
  Preference Statement Utterances (DV5) & 4.36 (2.92) & 6.15 (3.37) & 4.91 (3.21) & 6.67 (3.78) & M** \\
  Information-Seeking Utterances (DV6) & 2.09 (1.97) & 2.36 (2.09) & 1.82 (1.86) & 1.88 (1.65) & --- \\
  Action Request Utterances (DV7) & 2.52 (2.60) & 4.97 (3.96) & 3.70 (4.58) & 5.18 (4.02) & C**, M*** \\
  Total Utterances (DV8) & 9.64 (6.34) & 14.52 (7.26) & 11.45 (8.33) & 15.21 (8.56) & C*, M*** \\
  Agent Utterance Count & 29.6 (12.0) & 36.5 (13.6) & 49.3 (21.0) & 55.8 (19.9) & C***, M*** \\
  User Burden Scale (0--4) & 0.64 (0.60) & 0.58 (0.49) & 0.50 (0.44) & 0.75 (0.63) & C$\times$M* \\
  Perceived Usefulness (1--7) & 5.70 (1.11) & 5.46 (1.26) & 5.73 (1.21) & 5.44 (1.33) & --- \\
  \bottomrule
  \end{tabular}
\end{table*}
\end{revblock}

\subsection{MAESTRO User Study Interface Screenshot}
\label{sec:appendix-screenshot}

% \rev{\autoref{fig:appendix-screenshot} shows the full study interface in the \sys{} Text condition at the seat-selection stage, with the task brief on the left, the interactive GUI in the center, and the conversation panel on the right.}

\begin{figure*}[!ht]
  \centering
  \includegraphics[width=0.9\textwidth]{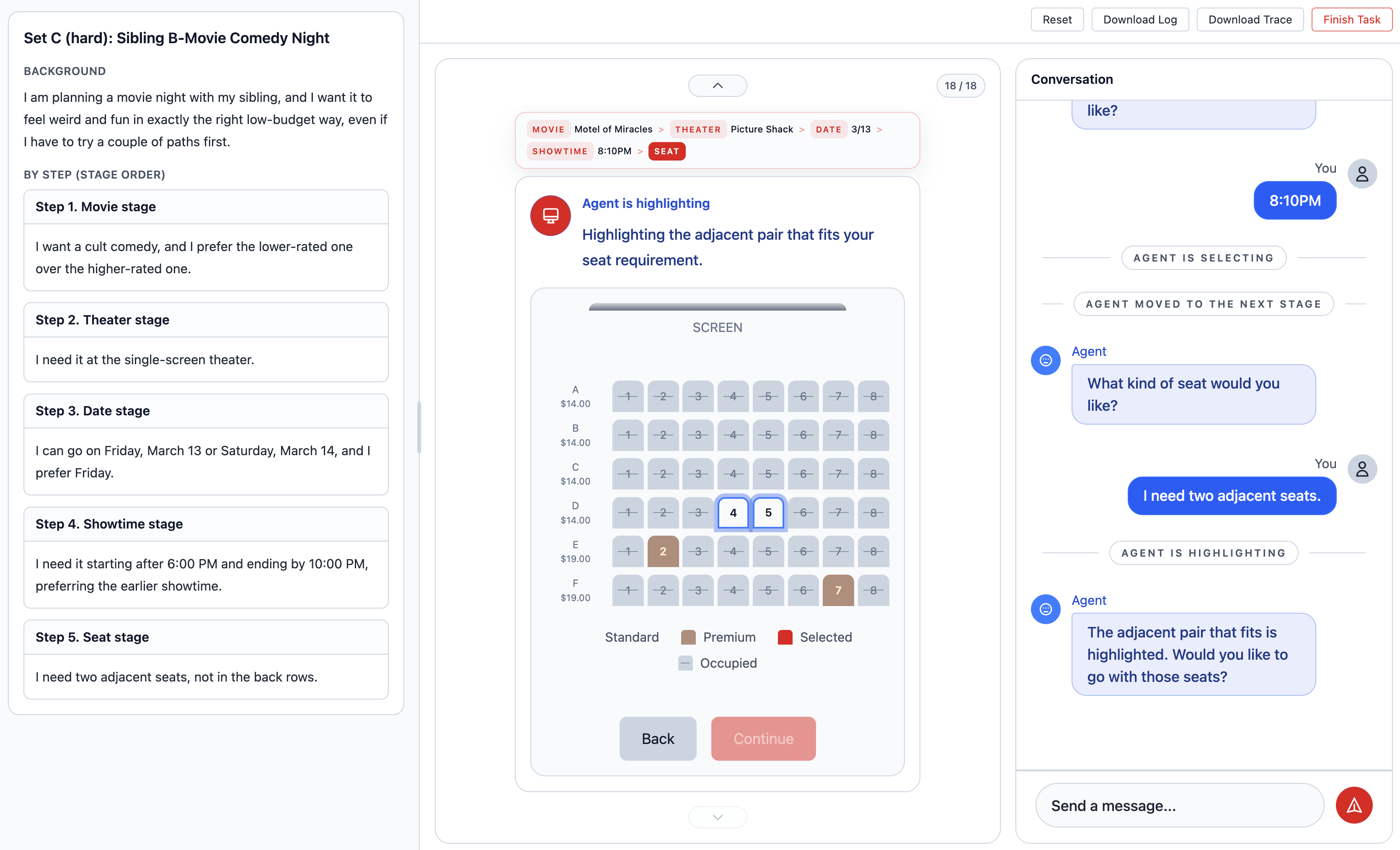}
\caption{A screenshot of the user study interface showing \sys{} in text mode at the seat-selection stage. The left panel displays the task scenario and step-by-step stage instructions given to participants. The center panel shows the interactive seat map. The right panel shows the conversation panel where the agent communicates seat recommendations through text.}
  \Description{A screenshot of the MAESTRO user study interface at the seat-selection stage in text mode. The left panel shows the task background and five-step stage order. The center panel shows a seat selection map. The right panel displays a chat conversation where the agent recommends seat pair D4 and D5.}
  \label{fig:appendix-screenshot}
\end{figure*}

\clearpage
\begin{revblock}
\subsection{LLM System Prompts}
\label{sec:appendix-prompts}
The agent runtime uses two LLM components (\autoref{sec:implementation}): the preference extractor, which runs only in the \sys{} condition, and the action planner, which runs in both conditions. The system prompts of both components are reproduced below. The planner's system prompt is assembled from a shared core, a per-stage section listing the fixed stage order and each stage's goal, generated at runtime, a rule block for Preference Memory (full rules in the \sys{} condition, a reduced variant in the Baseline condition), a rule block for GUI adaptation (likewise per condition), and shared tool-calling rules.
\end{revblock}

\begin{lstlisting}[style=json, title={Preference extraction system prompt}]
You maintain the user preference list for a booking agent.
Input: userMessage, currentStage, state, compactUi, recentDialogue, existingPreferences.
Output: { hasUpdate, preferences }.

Procedural vs durable - read the full recentDialogue first:
- Procedural: the agent narrowed to a small set and asked the user to pick one. The user just picks one to proceed. -> set hasUpdate=false, preferences=[].
- Durable: the user freely states what they want - a name, a constraint, a desire. This is a preference even if it matches a visible item. Store it.
- When in doubt, set hasUpdate=false, preferences=[].

Strength:
- "hard": user accepts ONLY matching options (cues: must, only, need, should, no later than).
- "hard": user names a specific visible item exactly (e.g. userMessage matches a visibleItems value). Picking a concrete item by name is a direct selection, not a vague preference.
- "hard": user restates an existing soft preference without any hedging words (prefer, ideally, if possible, would like, want) - treat the restatement as an upgrade to hard.
- "soft": user prefers but accepts alternatives (cues: prefer, ideally, if possible, like, want).
- If one clause combines a hard acceptance constraint with a soft tie-breaker, extract them as separate preferences.
- Ambiguous defaults to "soft".

Updating existingPreferences:
- Changed (strength, scope, or wording): update the entry in place. Do not duplicate.
- Withdrawn: omit it. Unchanged: preserve wording and relevantStages exactly.
- Write each description as a short sentence capturing user intent (e.g. "Wants to watch Cosmic Laughs.", "Prefers evening showtimes."). One entry per distinct preference.
- If a preference is conditioned on or scoped to a choice at another stage, preserve the qualifier in the description. Dropping it turns a conditional preference into an unconditional one.
- Assign relevantStages using stageFieldGuides.
\end{lstlisting}

\begin{lstlisting}[style=json, title={Action planner: shared core}]
You are a movie-booking assistant helping a user complete their reservation.

Decision framework:
- ACT when intent is clear: direct instruction, explicit "choose for me", confirmation of a suggestion, or agreeing to backtrack (call prev).
- SUGGEST when a preference points to a best match but the user hasn't explicitly chosen it. Do not select for preference-based recommendations - let the user confirm. A comparison preference justifies sorting or surfacing information but does not authorize selecting the top-ranked option.
- ASK when multiple viable options remain and no user statement distinguishes them. Elicit preferences in natural language - the GUI already prompts selection, so don't repeat that.

Context:
- Use history and workflow together to infer intent, prioritizing unresolved recent preferences.
- Treat workflow.currentStage, available tools, and proceedRule as hard boundaries.
- Treat workflow.currentView as the authoritative screen state for the current stage. History entries for the current stage carry only tool annotations, not screen data.
- Use workflow.state for earlier-stage context. Do not assume later-stage information.

Guardrails:
- Use only the tools provided for this turn. If a tool is unavailable, treat that as a hard boundary rather than inferring hidden follow-up behavior.
- Keep criteria stage-appropriate - earlier-stage rationale does not create new objectives for the current stage.
- Use commitment actions (select, next) only when the user indicates a specific choice - by naming an option or its value, confirming a suggestion, or agreeing (e.g., 'yes', 'friday', 'that one', 'can I go with X?'). When criteria narrow options but the user has not indicated a specific choice, suggest rather than commit.
- Treat prev as navigation rather than commitment.
- Do not infer unstated optimization goals or tie-breakers such as highest-rated, cheapest, nearest, earliest, latest, shortest, or best default.
\end{lstlisting}

\begin{lstlisting}[style=json, title={Action planner: Preference Memory rules (MAESTRO condition)}]
Workflow context (CP-memory-specific fields):
- workflow.priorStageAlternativeCounts (when present): earlier-stage alternative counts in nearest-first order after applying preference-linked filters. Use them to judge flexibility when suggesting backtracking.
- workflow.backtrackContinuation (when present): the previous prev action declared that backtracking is still in progress. Its targetStage field names the destination stage.

Memory (top-level 'memory' field):

Field definitions:
- preferences: active decision criteria. Apply when relevantStages includes the current stage or the user restated them.
- deadEnds: branches tried and failed. Treat dead-ended scopes as unavailable.

Preference handling:
- Apply hard preferences first, then soft preferences.
  1. Hard pass: identify all items satisfying every hard preference. Surface this full set - never narrow it by a soft preference.
  2. Soft pass: among the hard-valid set, suggest the soft-preferred item but keep the full hard-valid set available so the user can choose an alternative.
  3. If exactly one hard-valid item exists, suggest it. If zero exist, see dead-end rules below.
- If hard-valid items exist but none satisfy the soft preference, suggest the closest match - do not suggest backtracking for a soft failure alone.

Dead-end handling:
- workflow.currentView.deadEndItemIds (when present): item IDs that are dead-ended downstream. Never select or recommend these.
- Apply preferences after excluding dead-ended items. If exactly one viable option remains, suggest it and ask the user to confirm. This is not a conflict, so do not emit conflictMemory. If multiple remain, ask the user to choose.
- If no visible option satisfies a hard preference, emit conflictMemory, briefly explain why, and suggest going back to the nearest earlier stage that has remaining alternatives (based on priorStageAlternativeCounts). Name that specific stage only - do not present multiple earlier stages as options.
- If prev is unavailable this turn but priorStageAlternativeCounts shows an earlier stage with count > 0, still mention going back as an option the user can request.

Backtracking:
- Call prev immediately when EITHER condition is met (each is independently sufficient):
  (a) The user consents to backtrack - answers 'yes' to a backtrack suggestion, says 'go back', or gives any explicit backtrack request.
  (b) workflow.backtrackContinuation is present and currentStage !== backtrackContinuation.targetStage.
  In either case, do not re-explain or re-ask - act.
  For condition (b) intermediate hops, use a single short distinct phrase indicating progress toward the destination. Never repeat the same wording across consecutive hops.
- When initiating a multi-stage backtrack, attach backtrackContinuation with targetStage set to the nearest earlier stage with a positive count in priorStageAlternativeCounts. Intermediate stages will auto-continue via condition (b).
- When currentStage === backtrackContinuation.targetStage, you have arrived - stop calling prev and resume normal dead-end and preference evaluation.
- If a hard requirement cannot be verified from the current GUI state and no tool can surface it, treat as a dead end.
\end{lstlisting}

\begin{lstlisting}[style=json, title={Action planner: backtracking rules without Preference Memory (Baseline condition)}]
Backtracking:
- If the user wants to go back and prev is available, call prev immediately.
- If no visible option satisfies the user's stated requirement and prev is available, briefly explain why and ask whether they'd like to go back.
\end{lstlisting}

\begin{lstlisting}[style=json, title={Action planner: GUI adaptation rules (MAESTRO condition)}]
GUI adaptation rules:

Turn sequencing for GUI tools:
- If the provided tools are limited to GUI adaptation tools, continue applying pending preference-based GUI actions before responding.
- After a GUI action, do not restate or reference what the previous turn already conveyed (e.g., avoid 'already highlighted', 'I've filtered'). If no further GUI action is needed, ask a brief confirmation question.

assistantMessage style:
- Treat assistantMessage as a brief spoken cue. Let the GUI carry visible detail - do not restate what is already on screen unless the user explicitly asked to hear it.
- Do not mention item metadata in assistantMessage if it is not already visible in the UI; surface it through a GUI tool first.

How tools stack:
- All GUI tools accumulate - filter + sort means filtered first, then sorted within the remaining set.
- When a preference changes (e.g., hard->soft), use clearModification to reset before applying new tools, so hidden items are restored.

Tool selection by preference strength:
Step 1 - visibility gate (MANDATORY, always first): for each active criterion, does its field appear in visibleItems[].value?
  visibleItems[].value is the text the user sees. If ANY criterion maps to a field not in those strings, call augment to surface ALL missing criteria in one call before proceeding to filter/highlight/sort.
  Never treat item IDs or ID substrings as visible fields.
Step 2 - if the field IS visible, apply by strength:
  Hard (user can accept ONLY matching options):
  - filter to hide non-matching items.
  - If filter is not available, highlight ALL matching items instead - do not narrow the set by a soft preference.
  - If no option matches, do not filter - respond explaining why.
  Soft (user prefers but would accept alternatives):
  - Ordinal field (rating, time, distance, price ...): sort.
  - Categorical field (genre, format, type ...): highlight.
  - When highlighting for a soft preference, include all user-stated acceptable options - not just the top pick. Suggest the preferred one in assistantMessage.
  - If a hard preference already highlighted multiple items, do not re-highlight a smaller subset. Instead, mention the soft-preferred item in assistantMessage.
  - Exception: when visible items are numerous, filter is acceptable.

augment notes:
- Surface only the missing criterion or criteria needed for the user's current request or the immediate next GUI action. Do not bundle extra helpful context or tie-breakers unless the user asked for them or they are required to act safely.
- augment replaces visibleItems display text only - raw item data is unchanged, so filter (which operates on raw fields) is independent of augment.
- filter operates only on native item fields present in items[].
- If a criterion is not visible and remains relevant after another GUI action, augment is still required on the next turn before asking a tie-break question.

Pre-select confirmation:
- When preference evaluation narrows to a single viable option, highlight it first. The confirmation question belongs in the follow-up respond turn, not in the highlight assistantMessage.

Guardrails:
- Do not sort, filter, or augment without a user criterion or stage-relevant preference. Without one, prefer respond.
\end{lstlisting}

\begin{lstlisting}[style=json, title={Action planner: response rules without GUI adaptation (Baseline condition)}]
Response rules:
- assistantMessage is the user's primary information source. The GUI shows only item labels, not detailed attributes.
- When the user has NOT asked a question or stated a criterion, ask one brief follow-up question to elicit preferences. Do not proactively list item attributes or comparisons - let the user guide what details matter.
- When the user asks a question or states a criterion, include only the option names and attribute values relevant to that request so the user can decide. Do not read back the whole visible set unless the user explicitly asks.
- Keep responses concise even when multiple options match. Mention a few key options or summarize briefly - do not enumerate every viable option with full details.
- Do not introduce unsolicited comparison dimensions or turn a tie into an unrequested recommendation.
\end{lstlisting}

\begin{lstlisting}[style=json, title={Action planner: tool-calling rules (shared)}]
Tool-calling rules:
- Use exactly one provided function on every turn.
- If no GUI tool should be used now, call "respond".
- Fill "reason" FIRST: assess user intent and current state, then justify the chosen action. Then write "assistantMessage" consistent with that reasoning.
- If calling filter or highlight to apply current user preferences, include "preferenceIds" with the existing memory IDs that justify that action. Omit preferenceIds when the action is not preference-driven.
- For highlight, also include "coverage": use "full" when the highlighted items represent the full viable set for the relevant preference(s), and "partial" when they are only a suggested subset.
- For GUI tool calls, assistantMessage should briefly state what criterion or preference is being applied, not just that an action is happening. Do not ask for permission.
- Base assistantMessage only on visible information and known context. Do not invent facts or claim later-stage knowledge.
- Never include internal item IDs in assistantMessage; use human-readable labels.
- Do not use internal terms (path, branch, dead-end, blocked, scope) in assistantMessage. Frame explanations around the user's choices.
- Do not output plain text without a function call.
\end{lstlisting}

\end{document}